\documentstyle[12pt,aasms4,epsf]{report}
\def\ps@myheadings{\let\@mkboth\@gobbletwo
\def\@oddhead{{\sl\rightmark}\hfil \rm\thepage}}

\begin{document}
\newcounter{ctr}
\newcommand{\intinf}{\int_{-\infty}^{+\infty}}
\newcommand{\intinfh}{\int_{-h}^{+h}}
\newcommand{\Ostar}{\Omega_{*}}
\newcommand{\Oo}{\Omega_{0}}
\newcommand{\gm}{(\gamma^{*})}
\newcommand{\Ot}{\Omega_{2}}
\newcommand{\Ok}{\Omega_{k}}
\newcommand{\Dlnh}{\left({d \ln{h} \over d \ln{r}}\right)}
\newcommand{\Ddlnh}{{d \ln{h} \over d \ln{r}}}
\newcommand{\dlnh}{d \ln{h} / d \ln{r}}
\def\Ord#1{{\mathcal{O}}(\epsilon^{#1})}
\def\fulld#1#2{{d{#2}\over d{#1}}}
\def\gradd{\vec{\nabla}}
\def\divd#1{\grad\cdot{#1}}
\def\th#1{$#1$th}
\renewcommand{\thepart}{\Alph{part}}
\renewcommand{\thechapter}{\Roman{chapter}}
\renewcommand{\theequation}{{\mbox 
                                    {\arabic{section}.\arabic{equation}}}}
\def\epsffile#1{} 

\pagestyle{myheadings}

\bigskip

THREE-DIMENSIONAL STRUCTURE OF AN ALPHA
ACCRETION DISK


%
%
%

\centerline{W{\l}odzimierz Klu\'zniak and David Kita}

Physics Department, University of Wisconsin, Madison, WI 53706, USA

\vskip0.5truecm
\centerline{ABSTRACT}
\bigskip\par\noindent
An analytic solution is presented
to the three-dimensional problem of steady axisymmetric fluid flow
through an accretion disk.
The solution has been obtained through a systematic expansion in the small
parameter $\epsilon =\bar H/\bar R$ (the ratio of disk thickness to its
radial dimension) of the equations of viscous hydrodynamics.
The equation of state was assumed to be polytropic.
For all values $\alpha< 0.685$ of the viscosity parameter, we find 
significant backflow in the midplane of the disk occuring at all radii larger
than a certain value; however, in the inner regions
of the disk the fluid always flows toward the accreting object. The region
of backflow is separated from the region of inflow by a surface
flaring outwards from a circular locus of stagnation points 
situated in the midplane of the disk.

\bigskip\par\noindent
{\bf 1. Introduction}
\smallskip\par\noindent
Accretion flows occur in a variety  of astrophysical situations, often
they take the form of a disk (e.g.,
Frank, King \& Raine 1992). The first solutions for accretion
disk flows were constructed numerically by Prendergast \& Burbidge (1968),
while the first analytic solutions were obtained by Shakura \& Sunyaev (1973).
Since then, it has been the custom in analytic, but frequently also in
numerical, work to discuss
essentially one-dimensional solutions, i.e. to obtain the radial structure
of the disk by considering equations averaged over the thickness of the disk
and only then to obtain an approximate ``vertical'' structure by separately
considering equations describing hydrostatic equilibrium (and possibly
radiation transfer) in the direction perpendicular to the plane of the disk.
This is also true in discussions
 of quasi-spherical flows, such as in the
celebrated accretion-dominated flows (ADF) (e.g Narayan 1996 and references
therein)\footnote{In an important contribution
Narayan \& Yi 1995 go beyond the one-dimensional solutions by numerically
constructing axisymmetric ADF 
solutions which factorize the three-dimensional
equations, i.e., solutions of the type
$f(r, \theta)=R(r)\Theta(\theta)$. However, the solutions
we present in this paper are not factorizable.}. 

As already shown by Urpin (1984) in a remarkable paper,
 consideration of the vertical gradients
of the stress tensor leads to a solution in which the flow direction
in the midplane of the disk is opposite to that in the subsurface layers.
The flow cannot be properly described by its height-averaged value,
a point dramatically evident in the numerical work of Kley \& Lin 1992
who enforced spurious circulation flows in the meridional plane by adopting
``height-averaged'' boundary conditions at the edges of their computational
domain (nevertheless, they found and correctly identified an outflow
in the disk midplane, reminiscent of Urpin's solution). Several recent
numerical calculations (e.g. R\'o\.zyczka, Bodenheimer \& Bell 1994;
 Igumenshchev, Chen and Abramowicz 1995)
also exhibit flows which can best be described in terms of a tendency
for backflow to occur in the midplane of the disk. We believe this effect
is not thermal in origin, and to investigate the dynamics of the
phenomenon we solve analytically the
three-dimensional equations of disk accretion using a polytropic equation of
state for the fluid.

Urpin (1984) included thermal effects but made
the simplification of zero net angular momentum flow in the disk
(equivalently, his self-similar solution is valid asymptotically
for large radii). In our work we chose the opposite route---we neglect
thermal effects, but include the inner boundary condition---this allows
us to exhibit the global character of the solution. In particular,
we show how the backflow is fed by the inflowing fluid. In Section 2
we present the equations, in Section 3 we solve them. A discussion of the
results is begun in Section 4 and concluded, in Section 5, with a detailed
presentation of the velocity field in the disk.

\bigskip
\clearpage
\section{Disk equations. {\label{Ch2}}}
\thispagestyle{myheadings}
\bigskip

\subsection{Assumptions. {\label{assumptions}}}
\bigskip
We use cylindrical
coordinates $(r,\phi,z)$ centered on the accreting object
 and make the following standard assumptions:
\begin{list}{(\roman{ctr})}{\usecounter{ctr}}
\item
the gravitational force on a fluid element is characterized by the
Newtonian potential of a point mass,
\begin{equation}
\label{psi}
\psi(r,z) = - {GM_{*} \over \sqrt{r^{2}+z^{2}}},
\end{equation}
with $G$ the gravitational constant and $M_{*}$ the mass of the
central star;
\item
the structure of the disk is symmetric under reflection about the $(z=0)$
midplane;
\item
the disk is in a steady state $(\partial/\partial t=0)$;
\item
 the disk is axisymmetric $(\partial/\partial \phi=0)$,
hence all quantities will be expressed in terms of the coordinates
$(r,z)$;
\item
$\vert v_{\phi}\vert \gg \vert v_{r}\vert $;
\item The disk is geometrically thin, i.e. $\vert z\vert \ll r$;
\item
Viscous torques are a small perturbation in the radial ($r$) and
vertical ($z$) components of the equations of motion.
\end{list}
Assumption i) implies also that the disk is not self-gravitating.
The assumptions iii)--v) are consistent with the statement that
the accretion time scale is much greater than the Keplerian period.
Assumption vi) implies that
the rotational velocity is much greater than the local sound speed in
the outer parts of the disk, $v_{\phi} \gg c_{s}$, and that the radial
velocity is larger than the vertical one,
$\vert v_{r}\vert \ge \vert v_{z}\vert $.
Assumptions v)-vii) taken together signify 
that the disk is approximately in hydrostatic
equilibrium.

Throughout this paper we will also assume that
\par\noindent viii) the
equation of state for the disk is that of a
polytrope, i.e.
\begin{equation}
\label{Poly}
P= K \rho^{1+1/n},
\end{equation}
with $n$ and $K$ constant. Except in the Appendix we will take the 
polytropic index to be $n=3/2$.

For the inner boundary condition, we take vanishing of the viscous torque
at some radius $r_m$, corresponding to a maximum of the angular frequency,
 $\Omega_m$,  at the same radius.
This boundary condition, which introduces into the problem a natural
lengthscale, $r_+=\Omega_m^2r^4_{m}/(GM_*)$, is appropriate 
for black-hole disks and for stellar accretion disks
about stars which are spinning-up (whether magnetized or not).
For stars which are not spinning up, i.e. ones which transfer their angular
momentum to the disk (such as non-magnetized stars rotating
close to the equatorial mass-shedding limit and, possibly, for
X-ray pulsars [accreting neutron stars] in their spin-down phase), the solution
presented below is valid with the substitution
$1-\sqrt{r_+/r}\rightarrow 1+\sqrt{r_+/r}$. Thus, our solution
is universally valid for any thin, Keplerian accretion disk described
by a polytropic equation of state. We expect that the qualitative
features of our solution for the accretion flow will hold also for
other equations of state.

If the polytropic disk were in exact hydrostatic equilibrium, the angular
frequency $\Omega=v_\phi/r$ would be constant on cylinders and it would be
very easy to solve the equations of motion, at least far from the inner
boundary. In reality,
viscous terms (which are of order $\alpha^2$
in the alpha disk) break the hydrostatic equilibrium and cause the equations
of motion to form a system of nonlinear,
coupled, second order partial differential
equations which are rather challenging to solve (even numerically), but
which bring the reward of a solution whose salient features cannot be
described by height-integrated, i.e. ordinary, differential equations.

\subsection{Equations of motion. {\label{eqs_motion}}}
\bigskip

We use the generalized Navier Stokes equations, along with the
equation of continuity, to describe the accretion flow and to
represent viscous interactions:
\begin{equation}
\label{general_Navier}
\rho {d \vec{V} \over dt}+\rho (\vec{V}\cdot\vec{\nabla})\vec{V} =
-\vec{\nabla} P - \rho \vec{\nabla} \psi + \vec{\nabla}\cdot{\bf
\sigma},
\end{equation}
\begin{equation}
\label{general_mass_continuity}
{\partial \rho \over \partial t} + \vec{\nabla}\cdot(\rho\vec{V})=0,
\end{equation}
where $\rho$ is mass density, $P$ is pressure, $\vec{V}$ is the
velocity vector of a fluid element, and $\psi$ is the gravitational
potential.  The rank two viscous stress tensor, $\sigma$, is assumed
to have the following Cartesian components
(Landau \& Lifshitz \markcite{LL59} 1959):
\begin{equation}
\label{stress_tensor}
\sigma_{jk} = \eta \left [ {\partial V_{j} \over \partial x_{k} } +
{\partial V_{k} \over \partial x_{j} } - {2 \over
3}\delta_{jk}\vec{\nabla}\cdot\vec{V} \right] + \xi
\delta_{jk}\vec{\nabla}\cdot\vec{V},
\end{equation}
where $\xi$ is bulk viscosity and $\eta = \nu \rho$ is the dynamic
viscosity coefficient, both of which are functions of the coordinates.

With the assumptions described in $\S$~\ref{assumptions}, the
equations of motion in cylindrical coordinates become
\begin{equation}
\label{raw_Navier_radial}
v_{r}{\partial v_{r} \over \partial r} + v_{z}{\partial v_{r} \over
\partial z}-\Omega^{2}r=-{\partial \psi \over \partial r} - {1 \over
\rho}{\partial P \over \partial r} + {1 \over \rho}F_{r},
\end{equation}
\begin{equation}
\label{raw_Navier_angular}
\rho{v_{r} \over r^{2}}{\partial \over \partial r}(r^{2}\Omega) + \rho
v_{z}{\partial \Omega \over \partial z} = {1 \over r^{3}}{\partial
\over \partial r}\left(\eta r^{3}{\partial \Omega \over \partial
r}\right) + {\partial \over \partial z}\left(\eta {\partial \Omega
\over \partial z}\right),
\end{equation}
\begin{equation}
\label{raw_Navier_vertical}
v_{r}{\partial v_{z} \over \partial r} + v_{z}{\partial v_{z} \over
\partial z}= -{\partial \psi \over \partial z}-{1 \over \rho}{\partial
P \over \partial z} + {1 \over \rho}F_{z},
\end{equation}
\begin{equation}
\label{raw_mass_continuity}
{1 \over r}{\partial \over \partial r}(r \rho v_{r})+{\partial \over
\partial z}(\rho v_{z})=0,
\end{equation}
where
$\psi(r,z)$ is the gravitational potential given by
eq.~(\ref{psi}). $F_{r}$ and $F_{z}$ are respectively the $r$ and $z$
components of the divergence of the viscous stress tensor, i.e. the
viscous force, and are given by:
\begin{equation}
\label{F_r}
F_{r} = {2 \over r}{\partial \over \partial r}\left(\eta r {\partial
v_{r} \over \partial r}\right)- {2 \eta v_{r} \over r^{2}} + {\partial
\over \partial z}\left[\eta \left({\partial v_{r} \over \partial
z}+{\partial v_{z} \over \partial r}\right)\right]+ {\partial \over
\partial r}\left[\left(\xi-{2 \over 3}\eta
\right)\left(\vec{\nabla}\cdot \vec{V}\right) \right],
\end{equation}
\begin{equation}
\label{F_z}
F_{z} = {\partial \over \partial z}\left(2\eta {\partial v_{z} \over
\partial z}\right) + {1 \over r}{\partial \over \partial r}\left[\eta
r \left({\partial v_{r} \over \partial z}+{\partial v_{z} \over
\partial r}\right)\right]+ {\partial \over \partial
z}\left[\left(\xi-{2 \over 3}\eta \right)\left(\vec{\nabla}\cdot
\vec{V}\right) \right].
\end{equation}

\subsection{Constants of integration. {\label{constant_integration}}}
\bigskip

Vertical integration of
eq.~(\ref{raw_mass_continuity}), with the assumption of a steady
state, yields an expression of the conservation of mass flow through
cylinders. Usually this is written as {\mbox{$\dot{M}=-2\pi r \Sigma
\overline{v_{r}}$}} where $\dot{M}$ is the constant mass accretion rate
through any cylinder (and hence onto the star), $\Sigma$ is the
surface density in the disk, and $\overline{v_{r}}$ is an effective
(i.e. density-weighted, height-averaged) radial velocity. However, since we are
interested in the $z$ dependence of the radial velocity, $v_{r}$, we
choose to write this important equation as
\begin{equation}
\label{basic_mdot}
\dot{M} = -2\pi r \int_{-\infty}^{+\infty} \rho v_{r}\hspace*{1mm} dz
\hspace*{0.1cm}= \hspace*{0.15cm}{\mbox{constant}},
\end{equation}
where by convention $\dot{M} > 0$ for accretion, i.e. for
$\overline{v_{r}}<0$. The quantity $\dot{M}$ will serve as
an integral of the motion for our accretion flow.

Another constant is obtained if,
in the same spirit, we vertically integrate the angular momentum
equation (\ref{raw_Navier_angular}). If we first multiply both sides
by $r^{3}$ and integrate over $z$ from $-\infty$ to $+\infty$, we obtain:
\begin{equation}
\label{vert_1}
\int_{-\infty}^{+\infty} (r \rho v_{r}){\partial (r^{2}\Omega)\over
\partial r}dz -\int_{-\infty}^{+\infty} r^{3}\Omega{\partial (\rho
v_{z})\over \partial z} dz= {\partial \over \partial
r}\left[r^{3}\int_{-\infty}^{+\infty}\eta {\partial \Omega \over
\partial r} dz\right]+ \left. r^{3}\eta {\partial \Omega \over
\partial z}\right|_{-\infty}^{+\infty}
\end{equation}
where in deriving the second term on the left hand side we have
performed an integration by parts and set the boundary term to zero
since $\rho \rightarrow 0$ as $z \rightarrow \pm \infty$.

Using the equation of continuity (\ref{raw_mass_continuity}), we can
transform the entire left-hand side into $(\partial / \partial
r)\int_{-\infty}^{+\infty} r^{3} \rho v_{r}\Omega dz$. The last boundary
term involving $\partial \Omega / \partial z$
also vanishes because $\eta=\nu \rho \rightarrow 0$ as $\vert
z \vert \rightarrow \infty$ and, finally, integration over $r$
gives
\begin{equation}
\label{vert_2}
\dot{J}(r)- C =- 2\pi r^{3}\int_{-\infty}^{+\infty}\eta {\partial
\Omega \over \partial r} dz,
\end{equation}
where $-\dot{J}(r)$ is the advection rate of angular momentum through
a cylinder of radius $r$, $C$ is a constant of integration, and the
right-hand side is the net torque exerted by viscous interactions on
the same cylinder. Note that this equation is exact for any
azimuthally symmetric, steady flow in which no mass is exchanged through
the surface at infinity ($z=\pm\infty$), and in which no angular momentum
is carried radially by radiation (Klu\'zniak 1987).

Since we consider only cases when the accretion rate is never zero,
we can introduce another constant $j_+=C/\dot{M}$.
The torque vanishes when this new constant is equal to the height-averaged
specific angular momentum $\bar j$ (weighted with radial momentum
flux) or, correspondingly, when the height-averaged radial derivative
of the angular momentum
(weighted with dynamic viscosity) vanishes. That is, we can rewrite
eq.~(\ref{vert_2}) as:
\begin{equation}
\label{raw_vert_angular_final}
\dot{M}({\bar j} - j_{+}) = -2\pi r^{3}\left[\int_{-\infty}^{+\infty}\eta dz
\right] {\overline {d \Omega \over d r}},
\end{equation}
where $\dot{J}(r)=-2 \pi
r^{3}\left(\int_{-\infty}^{+\infty}  \rho v_{r}\Omega dz \right)=
\dot{M}{\bar j}$, etc.
If $\Omega$ is independent of $z$
(i.e. constant on cylinders), the usual form of
eq.~(\ref{raw_vert_angular_final}) is recovered by removing the bars.
Thus $j_{+}$ can be interpreted as the specific angular momentum at
the zero-torque radius, $r_{m}$.
For this reason, we now define a new
effective radius, $r_{+}$, at which the Keplerian specific angular
momentum is equal to $j_{+}$, i.e.
$\sqrt{GM_{*}r_{+}}=j_{+}=\Omega(r_{m})r_{m}^{2}$.
Note that in general the maximum value of
$\Omega$ is not equal to the corresponding Keplerian value,
$\Omega(r_m)\ne\Omega_k(r_m)$, and hence
we do {\underline{not}} expect $r_{+}$ to be the same as $r_{m}$.

\subsection{The polytropic sound speed. {\label{poly_sound_speed}}}
\bigskip

In the standard theory of thin accretion disks, the local sound speed
becomes of prime importance when modeling subsonic accretion. A clear
advantage of employing a barytropic equation of state is that it
reduces the number of variables by one. A polytropic equation of state also
greatly simplifies calculation of the local sound speed, i.e.
\begin{equation}
\label{sound_speed_def}
c_{s}^{2}={dP \over d\rho}=\left(1+{1 \over n}\right){P \over \rho}.
\end{equation} 
With the above relation we can rewrite the pressure gradients in
eqs.~(\ref{raw_Navier_radial}) \& (\ref{raw_Navier_vertical}) in terms
of $c_{s}$, giving the following elegant expressions:
\begin{equation}
\label{pressure_gradients}
{1 \over \rho}{\partial P \over \partial r} = n{\partial c_{s}^{2}
\over \partial r} \hspace*{0.5cm} ; \hspace*{0.5cm} {1 \over
\rho}{\partial P \over \partial z} = n{\partial c_{s}^{2} \over
\partial z}.
\end{equation}

Now it is easy to show a basic result concerning a fluid in
hydrostatic equilibrium in both the radial and vertical
directions. Here eqs.~(\ref{raw_Navier_radial}) \&
(\ref{raw_Navier_vertical}), without the inertial and viscous terms,
reduce to a simple form involving only $c_s$, $\Omega$, and $\psi$:
\begin{equation}
\label{radial_hydro_1}
-\Omega^{2} r = -{\partial \psi \over \partial r} - n{\partial
c_{s}^{2} \over \partial r },
\end{equation}
\begin{equation}
\label{vertical_hydro_1}
0=-{\partial \psi \over \partial z} -n{\partial c_{s}^{2} \over \partial z}.
\end{equation}
Taking $\partial / \partial r$ of eq.~(\ref{vertical_hydro_1}) and
$\partial / \partial z$ of eq.~(\ref{radial_hydro_1}), we obtain the
familiar result that $\partial \Omega / \partial z$=0, i.e. $\Omega$
is constant on cylinders for a polytrope in hydrostatic equilibrium
({\it{cf.}}~Tassoul \markcite{TAS78} 1978). Since in
eq.~(\ref{raw_Navier_angular}) the velocities are proportional to the
viscosity,
this already implies that in a barytropic disk of \underline{any thickness}
$\Omega$ is independent of
$z$ to leading order in a Taylor expansion in the (small) viscosity parameter
(except, possibly,
 when the specific angular momentum is constant, $j\equiv r^{2}\Omega=$
const.). We will perform a systematic expansion in a different small
parameter, the dimensionless disk thickness, but for subsonic flow
the same zeroth order result will be recovered.

\subsection{Scaling the equations of motion. {\label{scaling_perturb}}}
\bigskip

At this point it is of paramount importance that we scale all relevant
quantities by their corresponding characteristic values. This will
make the equations dimensionless and allow us to weigh the relative
significance of each term that appears. Following
Regev \markcite{REG83} (1983) we scale all velocities
($v_{z}$, $v_{r}$, and $c_{s}$) by the characteristic sound speed,
$\tilde{c_{s}}$, all radial distances by some characteristic radius
$\tilde R$ (e.g. $R_{*}$), and all vertical
distances by $\tilde{H}$, the typical vertical scale height in the
disk. We also represent $\Omega$ in units of
$(GM_{*}/\tilde R^{3})^{1/2}\equiv \Omega_{k*}$, the Keplerian angular
velocity at the characteristic radius, and $\rho$ in terms of a typical value
$\tilde{\rho}$. Similarly, we scale the pressure by
$\tilde{P}=\tilde{\rho}\hspace*{1.0mm} \tilde{c_{s}}^{2}$,
the kinematic viscosity by $\tilde{\nu}=\tilde{c_{s}}
\tilde{H}$, and the dynamic and bulk viscosity coefficient by
$\tilde \zeta=\tilde{\eta}=\tilde{\nu} \hspace*{1.0mm} \tilde{\rho}$.

To apply a perturbative expansion technique to each equation we define
an expansion parameter, $\epsilon$. Since we are interested in
geometrically thin disks we choose
\begin{equation}
\label{epsilon_def}
\epsilon = {\tilde{H} \over \tilde R} = {\tilde{c_{s}} \over
\Omega_{k*}\tilde R} \ll 1,
\end{equation} 
where we have used $\tilde{c_{s}}=\tilde{H} \Omega_{k*}$, in
agreement with the standard result from thin disk theory that $H \sim
c_{s}/\Omega_{k}$. In effect, $\epsilon$ is a parameter which measures
the relative ``thinness'' of the disk.

Denoting the scaled forms of $v_{r}$ and $v_{z}$ by $u$ and $v$
respectively, we obtain the following set of non-dimensional
equations:
\begin{eqnarray}
\label{scaled_radial_1}
\nonumber \epsilon^{2}u{\partial u \over \partial r}+ \epsilon
v{\partial u\over \partial z}- \Omega^{2}r = - {1 \over
r^{2}}\left[1+\epsilon^{2}\left({z\over r}\right)^{2}\right]^{-3/2}
-\epsilon^{2}\left(n {\partial c_{s}^{2} \over \partial r}\right)-
\epsilon^{3}\left({2\eta u \over \rho r^{2}} \right)\\ +
{\epsilon^{3}\over \rho r} {\partial \over \partial r}\left(2 \eta r
{\partial u \over \partial r} \right) + {\epsilon \over \rho}{\partial
\over \partial z}\left(\eta{\partial u \over \partial z}\right)+
{\epsilon^{2}\over \rho}{\partial \over \partial z}\left(\eta{\partial
v \over \partial r}\right)\\ \nonumber +{\epsilon^{3}\over
\rho}{\partial \over \partial r}\left[\left(\xi-{2 \over 3}\eta
\right)\left({1 \over r}{\partial \over \partial r}(ru)
\right)\right]+{\epsilon^{2}\over \rho}{\partial \over \partial
r}\left[\left(\xi-{2 \over 3}\eta \right){\partial v\over \partial
z}\right],
\end{eqnarray}
\begin{eqnarray}
\label{scaled_angular_1}
\epsilon {\rho u \over r^{2}}\left[{\partial \over \partial r} (r^{2}
\Omega) \right] + \rho v{\partial \Omega \over \partial z}=
\epsilon^{2}\left[{1 \over r^{3}} {\partial \over \partial
r}\left(\eta r^{3} {\partial \Omega \over \partial r}\right) \right] +
{\partial \over \partial z}\left(\eta {\partial \Omega \over \partial
z} \right),
\end{eqnarray}
\begin{eqnarray}
\label{scaled_vertical_1}
\nonumber \epsilon u {\partial v \over \partial r} + v{\partial v
\over \partial z}= - {z \over r^{3}}\left[1+\epsilon^{2}\left({z\over
r}\right)^{2}\right]^{-3/2}-n {\partial c_{s}^{2} \over \partial z}+
{2 \over \rho}{\partial \over \partial z}\left(\eta {\partial v \over
\partial z} \right)\\ + {\epsilon^{2} \over \rho r} {\partial \over
\partial r}\left(\eta r {\partial v \over \partial r}
\right)+{\epsilon \over \rho}{\partial \over \partial
z}\left[\left(\xi-{2 \over 3}\eta \right)\left({1 \over r}{\partial
\over \partial r}(ru) \right)\right]\\ \nonumber +{ 1 \over
\rho}{\partial \over \partial z}\left[\left(\xi-{2 \over 3}\eta
\right){\partial v\over \partial z}\right] + {\epsilon \over \rho r}
{\partial \over \partial r}\left(\eta r {\partial u \over \partial z}
\right),
\end{eqnarray}
\begin{equation}
\label{scaled_mass_1}
{\epsilon \over r}{\partial \over \partial r}(r \rho u) + {\partial
\over \partial z}(\rho v) = 0,
\end{equation}
where we have used eq.~(\ref{pressure_gradients}) in rewriting the
pressure gradients. Here eqs.~(\ref{scaled_radial_1}),
(\ref{scaled_angular_1}), and (\ref{scaled_vertical_1}) are the scaled
radial, angular, and vertical momentum equations respectively and
eq.~(\ref{scaled_mass_1}) is the scaled form of the continuity
equation. Armed with the knowledge that $\epsilon \ll 1$ for a thin
disk, we make
eqs.~(\ref{scaled_radial_1})-(\ref{scaled_mass_1}) the foundation of
our analysis and proceed to perturbatively expand all dynamical
quantities in powers of $\epsilon$.
We will find $u\sim\epsilon$, $v\sim\epsilon^2$, i.e.,
$v_r={\mathcal O}(\epsilon^2)v_\phi$ and $v_z=
{\mathcal O}(\epsilon^3)v_\phi$; therefore
the divergence terms $\nabla\cdot\vec V$ in eq.~(\ref{stress_tensor})
contribute at order not lower than $\epsilon^3$ to
eqs.~(\ref{raw_Navier_radial}) and (\ref{raw_Navier_vertical})---this
formally justifies their frequent neglect.

\section{Solution for the vertical structure by perturbative expansion in
$\epsilon$. {\label{Ch3}}}
\thispagestyle{myheadings}
\bigskip

\subsection{Power series in $\epsilon=\tilde H/\tilde R$. {\label{prelims}}}
\bigskip

We expand all variables in powers of
$\epsilon$ and will evaluate
eqs.~(\ref{scaled_radial_1})-(\ref{scaled_mass_1}) at various
orders. We let
\begin{equation}
\label{omega_exp}
(\Omega/ \Omega_{k*}) = \Omega_{0} + \epsilon
\Omega_{1} + \epsilon^{2} \Omega_{2} + ..., 
\end{equation}
\begin{equation}
\label{u_exp}
u=(v_{r}/\tilde{c_{s}})= u_{0} + \epsilon u_{1} + \epsilon^{2}
u_{2} + ..., 
\end{equation}
\begin{equation}
\label{v_exp}
v=(v_{z}/\tilde{c_{s}})= v_{0} + \epsilon v_{1} +
\epsilon^{2} v_{2} + ..., 
\end{equation}
as well as
$(c_{s}/\tilde{c_{s}})
= c_{s0} + \epsilon c_{s1} + \epsilon^{2} c_{s2} +$ ...,
and $(\rho /\tilde{\rho}) = \rho_{0} +
\epsilon \rho_{1} + \epsilon^{2} \rho_{2} +$ ..., with the assumption
that the dimensionless vertical scale height of the disk,
$h(r)=H(r)/\tilde{H}$, is of order unity, 
$h\sim{\mathcal{O}}(1)$. All other variables like $P$, $\eta$,
$\nu$, \& $\dot{M}$ can be expressed in terms of these six fundamental
quantities{\footnote{In reality only five of these are independent
for a polytrope
since eq.~(\ref{sound_speed_def}) gives $c_{s}$ in terms of $\rho$ (or
vice-versa).}}. Our objective then is to calculate, order by order in
$\epsilon$, the functional dependence of $\Omega$, $u$, $v$, $c_{s}$,
$\rho$, and $h$ on the coordinates $r$ and $z$ alone.

Assumption ii) of $\S$~\ref{assumptions}, regarding
reflection symmetry about the ($z=0$) midplane, implies that physical
quantities such as $\Omega$, $\rho$, $P$, $\eta$, $u$, and $c_{s}$ are
even functions of $z$, while $v$ is odd under reflections through the
equatorial plane. When we expand an even/odd function (e.g. $\Omega$)
in powers of $\epsilon \ll 1$, we require each term in the expansion
{\mbox{(e.g. $\Omega_{i}$; $i=0,1,2,$...)}} to be independently
even/odd.

\subsection{Viscosity-independent
zeroth order results for the vertical structure. {\label{generic_zero}}}

Examination of eq.~(\ref{scaled_radial_1}) at zeroth order
immediately gives
\begin{equation}
\label{omega_kep}
\Omega_{0} = r^{-3/2},
\end{equation}
i.e. $\Omega_{0}$ is equal to the Keplerian value at the midplane,
$\Omega_k\equiv v_{\phi k}/r\equiv\sqrt{GM_*/r^3}$ in conventional units.
Though eq.~(\ref{omega_kep}) is consistent with the assumption of a
rotationally supported disk, it cannot satisfy the inner boundary
condition of $\Omega(r_{*})=\Omega_{*}$ whenever the star rotates
below its Keplerian value at the stellar radius, nor the more general
zero-torque boundary condition. Clearly, our perturbative solution
will be invalid in the limit $r\rightarrow r_+$. This is because
implicit in the scaling of
$\S$~\ref{scaling_perturb} has been the assumption that $\partial/
\partial r \sim \epsilon (\partial/ \partial z)$. This approximation,
however, is known to be patently false in the inner transition region
between the central star and the Keplerian portion of the disk.

Eq.~(\ref{scaled_mass_1}), the equation of continuity, fixes $v$
at zeroth order to be zero everywhere, $v_{0} =0$.
To see this more clearly, observe that
$\partial (\rho_{0} v_{0})/ \partial z =0$ and so $ \rho_{0} v_{0}$,
the lowest order vertical component of the mass flow, is a function of
$r$ only. However, since $v$ is odd with respect to the $z$
coordinate, we know $\rho_{0} v_{0}=0$ for all points on the midplane
($z=0$) and, not being a function of $z$, this product must then
vanish everywhere. Clearly $\rho_{0} \ne 0$ and thus $v_{0}=0$ at all
points in the rotationally supported disk.

Moving on to first order in $\epsilon$ for the angular velocity, we
see that because $\Omega$ is even with respect to reflections through
the midplane, the first order correction to the angular velocity vanishes,
$\Omega_{1}=0$.
This result for $\Omega_{1}$ also has direct impact on the fluid
velocities, since eqs.~{\mbox{(\ref{scaled_angular_1})}} \&
(\ref{scaled_mass_1}) at order $\epsilon$ now give
$u_{0}=v_{1}=0$. Using this result for $u$ and $v$, we then find
that the only surviving term of order $\epsilon$ in
eq.~(\ref{scaled_vertical_1}) involves the first order correction to
the square of the sound speed, and thus $c_{s1}^{2}=0$, and hence
$\rho_{1}=P_{1}=0$, which is consistent with symmetry arguments for
these three quantities. With this information,
eq.~(\ref{scaled_vertical_1}) simply becomes the standard equation of
vertical hydrostatic equilibrium:
\begin{equation}
\label{vert_hydro_0}
{1 \over \rho_{0}}{\partial P_{0} \over \partial z}=n{\partial
(c_{s0}^{2}) \over \partial z}= -{z \over r^{3}}.
\end{equation} 

We can solve eq.~(\ref{vert_hydro_0}), i.e. the vertical momentum
equation up to corrections of order $\epsilon^{2}$, and hence
find
 $c_{s0}$, $\rho_{0}$, and $P_{0}$.
In the case of polytropic index $n=3/2$ we obtain the
following relations (H\={o}shi\markcite{HO77}, 1977):
\begin{equation}
\label{sound_speed_0}
c_{s0}(r,z) = \sqrt{{h^{2}-z^{2} \over 3r^3}},
\end{equation}
\begin{equation}
\label{rho_0}
\rho_{0}(r,z) = {\left(h^{2}-z^{2}\over 5 r^{3}\right)}^{3/2},
\end{equation}
\begin{equation}
\label{P_0}
P_{0}(r,z) = \rho_{0}^{5/3} = {\left(h^{2}-z^{2}\over 5
r^{3}\right)}^{5/2}.
\end{equation}
Eqs.~(\ref{rho_0})-(\ref{P_0}) show that $h(r)$ is now the height at
which $\rho=0$ (and hence $P=0$), implying that $h$ is the true
semi-thickness of the disk, though its functional dependence on $r$ is
still undetermined. The surface density, $\Sigma(r)$, can also be
derived in terms of $h$ to lowest order in $\epsilon$:
\begin{equation}
\label{sigma_0}
\Sigma_{0}(r) = \int_{-h}^{+h} \rho_{0}\hspace*{1.0mm}dz = {3\pi \over
40\sqrt{5}}{\left(h^{4}\over r^{9/2}\right)},
\end{equation}
where we have replaced the integration limits of $\pm\infty$ by $\pm
h$, since by eq.~(\ref{rho_0}) the polytropic disk terminates at
$z=\pm h(r)$.

It is now possible to put the integrated mass continuity equation
(\ref{basic_mdot}) into dimensionless form. Scaling as before and
using the fact that to lowest non-vanishing order in $\epsilon$,
$u=\epsilon u_{1}+...$ and $\rho=\rho_{0}+...$, we can write
eq.~(\ref{basic_mdot}) to lowest order in $\epsilon$:
\begin{equation}
\label{scaled_vert_angular_1}
\epsilon \hspace*{0.5mm}\dot{m}= - r \int_{-h}^{+h} \rho
u\hspace*{0.5mm} dz \sim -\epsilon\hspace*{0.5mm} r \int_{-h}^{+h}
\rho_{0} u_{1}\hspace*{0.5mm} dz,
\end{equation} 
where $\dot{m}={\dot{M}}/(2\pi \tilde{\rho}{\hspace*{0.25mm}}
\tilde{c_{s}} \tilde{H}^{2})\sim{\mathcal{O}}(1)$, is a
dimensionless constant. For an adiabatic index of $n=3/2$,
eq.~(\ref{Poly}) gives $\tilde{\rho}\sim
K^{-3/2}\tilde{c_{s}}^{3}$ and, since $\tilde{H} \sim
\tilde{c_{s}}/ \Omega_{k*}$, we
find $\dot{m}\sim \dot{M}/\tilde{c_{s}}^{6}$. Because $\dot{m}$ is
independent of $\epsilon$ for $u_{1} \ne 0$, this implies that the
unscaled mass accretion rate must scale as $\dot{M}\sim
{\mathcal{O}}(\epsilon^{6})$, and hence $\dot{M}$ depends sensitively
on the relative ``thinness'' of the disk. As we shall soon see, the
scaled eq.~(\ref{scaled_vert_angular_1}) is of prime importance in
determining the vertical structure and must therefore be included with
eqs.~(\ref{scaled_radial_1})-(\ref{scaled_mass_1}).

Up to this point, all the results obtained in this section, including
eqs.~(\ref{omega_kep}) and (\ref{sound_speed_0})-(\ref{sigma_0}), are
common to the outer regions ($r>>r_+$)
of any standard thin disk with polytropic
equation of state, for any viscosity prescription.
 However, all of these expressions depend intrinsically on
the vertical scale height, $h(r)$, which cannot be determined as an
explicit function of $r$ without a form for $\eta(r,z)$ being first
specified. We
cannot obtain solutions for the lowest non-vanishing orders in
$\epsilon$ of $v_{r}$ and $v_{z}$, nor can we evaluate the first
nonzero correction to $\Omega$ without specifying the viscosity prescription.

\subsection{Lowest-order results using the standard $\alpha$-disk prescription.
{\label{more_lowest}}}
\bigskip

Let us continue all calculations under the assumption that
the viscosity is given by the $\alpha$-disk formulation.
In view of the large uncertainty in modeling the kinematic viscosity,
authors in the past have generally neglected any $z$ dependence
for $\nu$. Laboratory studies of turbulent jets undergoing
free expansion lend some support to this hypothesis ({\it{cf.}}~Urpin
\markcite{URP84a} 1984a, Monin \& Yaglom \markcite{MY65} 1965).
However, we find
this to be unacceptable when solving for $v_{r}(r,z)$ and $v_{z}(r,z)$
in a polytropic disk as it leads to divergent expressions at the
surface of the disk, $v(r,\pm h)\rightarrow\infty$ (Kita 1995).
We choose then to modify the alpha prescription
by directly incorporating a form of $z$ dependence into the kinematic
viscosity. But first, to better demonstrate that the zeroth order
results are hardly affected by the choice of the $z$ dependence in the
kinematic viscosity,
let us write down the result for the height of the
(polytropic) standard  alpha disk, where
$\partial\nu/\partial z=0$ and
\begin{equation}
\label{alpha}
\nu=\alpha c_{s}H.
\end{equation}

If $c_{s}$ taken to be the zeroth order equatorial sound speed,
$\bar c_{s0}(r)\equiv c_{s0}(r,0)$, one obtains the following zeroth
order  expressions for
the kinematic and dynamic viscosity coefficients:
\begin{equation}
\label{nu_0}
\bar \nu_{0}(r)={\alpha \over \sqrt{3}}\left({h^{2} \over r^{3/2} } \right),
\end{equation}
\begin{equation}
\label{eta_0}
\bar \eta_{0}(r,z)= \bar \nu_{0}\rho_{0}= {\alpha \over 5\sqrt{15}}
\left[{h^{2}\left(h^{2}-z^{2}\right)^{3/2} \over
r^{6}}\right].
\end{equation}
Notice that with this standard $\alpha$-disk viscosity law, $\nu_{0}$
depends solely on $r$ and, therefore, $\eta_{0}$ inherits its
$z$ dependence entirely from the density, $\rho_{0}(r,z)$, as given by
eq.~(\ref{rho_0}).

Now the disk semi-thickness, $h(r)$, can be determined as a function
of radius. To do this we observe that through lowest order in
$\epsilon$,
eq. (\ref{raw_vert_angular_final}) reads:
\begin{equation}
\label{scaled_angular_integrated}
\dot{m} (j_{0}-j_{+}) = -r^{3}\left[\int_{-h}^{+h} \eta_{0}dz\right]
{d \Omega_{0}\over d r},
\end{equation}
where $\dot{m}$ is the scaled constant from
eq.~(\ref{scaled_vert_angular_1}), $j_{0}=r^{1/2}$ is the (zeroth
order) Keplerian specific angular momentum, and $j_{+}=r_{+}^{1/2}$ is
the integration constant that arises from the no-torque boundary
condition{\footnote{If for some reason $\Omega_{0}$ was a function of
both $r$ and $z$ we would need to use eq.~(\ref{vert_2}).}} that
accompanies eq.~(\ref{raw_vert_angular_final}).
Using eq.~(\ref{eta_0}) for $\eta_{0}$ and integrating over $z$, we
are left with the following simple algebraic equation for the
dimensionless disk height, $h(r)$:
\begin{equation}
\label{dimensionless_h}
{h(r) \over r}= \bar \lambda_0\left(1- \sqrt{{r_{+} \over r}}\right)^{1/6}
\hspace*{1.0cm} {\mbox{with}} \hspace*{1.0cm} \bar \lambda_0=\left[{\dot{m}
\over \alpha} \left({80 \over 3\pi}\sqrt{{5 \over 3}}\right)
\right]^{1/6}.
\end{equation}
This result implies that $h(r)/r \rightarrow \bar \lambda_0=$constant as $r
\rightarrow \infty$, and that the disk remains thin (i.e. $H(r)/r \sim
\epsilon$) for all radii, provided, of course, that $\alpha$ is not
too small. In addition, we see that as $r \rightarrow r_{+}$, $h
\rightarrow 0$. This is a consequence of our using $j_{0}$ and $d
\Omega_{0}/ d r$ in eq.~(\ref{scaled_angular_integrated}), despite the
requirement that $d \Omega / d r \rightarrow 0$ as $j$ (not $j_{0}$)
$\rightarrow j_{+}$. As we will see in Section 4, this is in fact
a signal that our use of eq. (\ref{scaled_angular_integrated}) to
determine $h$ is not appropriate in the neighborhood of $r_{+}$.

It should be pointed out that the results we have so far
obtained in this and the previous subsection are well known
({\it{cf.}}~Shakura \& Sunyaev \markcite{SS73} 1973, H\={o}shi
\markcite{HO77} 1977, Paczy\'{n}ski \markcite{PAC91} 1991). Novel
developments only become apparent when
eqs.~(\ref{scaled_radial_1})-(\ref{scaled_mass_1}) are solved to
higher order. But, as already remarked,
 eqs.~(\ref{nu_0}) and (\ref{eta_0}) cannot be used to
consistently extend the results obtained so far to higher order, it is
first necessary to slightly modify the viscosity prescription.

\subsection{Lowest order results with height-dependent kinematic viscosity.
 {\label{mod_visc}}}
\bigskip
In the original paper by Shakura \&
Sunyaev \markcite{SS73} (1973) the dominant component of the
viscous stress tensor is presumed to have the following form:
\begin{equation}
\label{alpha_basic}
\sigma_{r \phi} \sim \eta r (\partial \Omega / \partial r) \sim
-\alpha P.
\end{equation}
Since we know that in the outer disk $r(\partial
\Omega / \partial r)\sim -\Omega_{k}$, we use eq.~(\ref{alpha_basic})
as a guide to make the following assumption regarding the
$z$ dependence of the viscosity:
\begin{equation}
\label{mod_eta}
\nu(r,z) = {\alpha \over
\Omega_{k}}\left[{c_{s}^{2}(r,z) \over (1+1/n)}\right],
\end{equation}
consistent with $\eta(r,z)\sim {\alpha P(r,z)/\Omega_{k}}$
where $P \sim \rho c_{s}^{2}$ for a polytrope. 
Our new expression for $\nu(r,z)$ reduces to the original
formulation of eq.~(\ref{alpha}) in the midplane of the disk
($z=0$).
We justify our choice of $\nu$ by noting that if the turbulent speed,
$v_{turb}$, is bounded from above by the local
sound speed, we must expect $v_{turb}$ to vary considerably
with height, since for a polytrope $c_{s}\rightarrow 0$ as $z
\rightarrow \pm h$. 

By using eqs.~(\ref{sound_speed_0})-(\ref{P_0}) \&
(\ref{mod_eta}), we derive new forms for the zeroth order kinematic
and dynamic viscosity coefficients:
\begin{equation}
\label{nu_0_2}
\nu_{0}(r,z)={2 \alpha \over 15}\left[{h^{2}-z^{2} \over
r^{3/2}}\right],
\end{equation}
\begin{equation}
\label{eta_0_2}
\eta_{0}(r,z)=\nu_{0}(r,z)\rho_{0}(r,z)= {2 \alpha \over
75\sqrt{5}}\left[{\left(h^{2}-z^{2}\right)^{5/2} \over r^{6}}\right].
\end{equation}
Comparison of eq.~(\ref{eta_0_2}) with eq.~(\ref{eta_0}) shows that
$\eta_{0}$ is now one power higher in $(h^{2}-z^{2})$ and, as shown by Kita
1995, it is this difference that is
the key to suppressing the divergence of $v_{r}$ and $v_{z}$ at the
disk surface.


We must first examine the impact that
eqs.~(\ref{nu_0_2})-(\ref{eta_0_2}) may have on all other previously
derived quantities.
Since none of the results in $\S$~\ref{generic_zero} depend in any way
on $\nu$ or $\eta$, we know they are unaffected. The only change is
in the value of the height of the disk (but not its functional form), which
is increased by a factor $3^{1/4}$:
\begin{equation}
\label{lambda_true}
{h(r) \over r}= \lambda\left(1- \sqrt{{r_{+} \over r}}\right)^{1/6}
\hspace*{0.2cm} {\mbox{with}} \hspace*{0.3cm}
\lambda=\left[{\dot{m} \over \alpha} \left({16 (5)^{3/2}\over
\pi}\right) \right]^{1/6} \approx 1.96 \left({\dot{m} \over \alpha
}\right)^{1/6}.
\end{equation}

\subsection{Second order disk equations. {\label{real_equations}}}
\bigskip

To explore the differential rotation with respect to $z$
and the nature of the velocity vector field in the accretion disk, we now
consider only terms of ${\mathcal{O}}(\epsilon^{2})$
in eqs.~(\ref{scaled_radial_1})-(\ref{scaled_angular_1}).
Bearing in
mind that $\Omega_{0}=r^{-3/2}$, $\Omega_{1}=0$, and
$u_{0}=v_{0}=v_{1}=0$, we discover the following equations for
$\Omega_{2}$, $u_{1}$, and $v_{2}$:
\begin{equation}
\label{real_radial}
-2\Omega_{0} \Omega_{2} r = {3 \over 2}{z^{2} \over r^{4}}- {3 \over
2}{\partial c_{s0}^{2} \over \partial r} + {1 \over \rho_{0}}{\partial
\over \partial z}\left(\eta_{0} {\partial u_{1} \over \partial z}
\right),
\end{equation}
\begin{equation}
\label{real_angular}
r\rho_{0}u_{1}{d \left(r^{2} \Omega_{0} \right)\over d r}= {\partial
\over \partial r}\left(\eta_{0}r^{3}{d \Omega_{0} \over d r} \right) +
r^{3}{\partial \over \partial z}\left(\eta_{0} {\partial \Omega_{2}
\over \partial z} \right),
\end{equation}
\begin{equation}
\label{simple_continuity}
{1 \over r} {\partial \over \partial r}(r \rho_{0} u_{1}) + {\partial
\over \partial z}(\rho_{0} v_{2}) = 0.
\end{equation}
Here, $c_{s0}$, $\rho_{0}$ and $\eta_0$ are given by
eqs.~(\ref{sound_speed_0}), (\ref{nu_0_2}) and (\ref{eta_0_2}), and
$h$ is known up to the integration constant $r_{+}$.
 Unfortunately, the two viscous
terms at the end of eqs. (\ref{real_radial}) \& (\ref{real_angular})
complicate things by coupling the two
equations together. To simplify the equations,
most authors assume {\it a priori} that $\Omega$
and $v_{r}$ in the outer disk are both functions of only $r$. As we will
show, this is justified only in the limit $\alpha<<\epsilon$.
We prefer to keep all terms so that we can obtain a solution valid
through order $\epsilon^3$ which is consistent for all values of $\alpha$.

\subsection{Complete analytical solution for $\Omega_{2}$, $u_{1}$, and $v_{2}$.
 {\label{real_results}}}
\bigskip

To solve for $\Omega_{2}$ and $u_{1}$, we will make
the ansatz that \linebreak \noindent
$u_{1}(r,z)=f_{1}(r)(h^{2}-z^{2})+f_{2}(r)$, where $f_{1}(r)$ and
$f_{2}(r)$ are as yet undetermined\linebreak \noindent functions of
$r$. A heuristic justification for this choice is that if the equations
are decoupled by neglecting the $z$ derivatives,
as in the olden approach common in the literature, the
solution for the lowest order corrections to Keplerian motion are
\begin{equation}
\label{novisc_omega}
{\Omega_{2}|^{old}(r) \over \Omega_{0}}= -{3 \over 4}\left({h \over
r}\right)^{2}\left[1 - {2 \over 3}\left({d \ln{h} \over d
\ln{r}}\right) \right],
\end{equation}
and
$u_{1}|^{old}(r,z)=g_{1}(r)(h^{2}-z^{2})+g_{2}(r)$, where
\begin{equation}
\label{gs}
g_{1}(r)= \left({11 \over 5 }\right) {\alpha \over r^{5/2}}
\hspace*{1.0cm}{\mbox{{and}}} \hspace*{1.0cm} g_{2}(r)=-2\alpha
\left({h^{2}\over r^{5/2}}\right)\left({d \ln{h} \over d
\ln{r}}\right).
\end{equation}
Given the nature of the coupling term, $\eta_{0} (\partial u_{1}/
\partial z)$, in eq.~(\ref{real_radial}), it is now possible to
formulate $\Omega_{2}$ in terms of $f_{1}(r)$ and $f_{2}(r)$, i.e.
\begin{equation}
\label{omega_old}
\Omega_{2}(r,z)= \Omega_{2}|^{old}(r) + {2 \over 15}
\alpha\left({f_{1}(r) \over r}\right)(h^{2}-6z^{2}).
\end{equation}

We then solve for $f_{1}(r)$ and $f_{2}(r)$ by substituting
eq.~(\ref{omega_old}) directly into the term involving
$\eta_{0}(\partial \Omega_{2}/\partial z)$ that appears in
eq.~(\ref{real_angular}). This results in the following forms for
the radial functions:
\begin{equation}
\label{fs}
f_{1}(r)= {g_{1}(r) \over \left(1 + {64 \over 25}\alpha^{2}
\right)}\hspace*{1.0cm} {\mbox{{and}}} \hspace*{1.0cm}
f_{2}(r)=\left({32 \over 15}\alpha^{2} \right)\left({g_{1}(r)
h^{2}\over 1+{64 \over 25} \alpha^{2}} \right)+ g_{2}(r),
\end{equation}
where the functions $g_{1}(r)$ and $g_{2}(r)$ are defined by
eq.~(\ref{gs}). In this way we finally obtain complete
solutions for $\Omega_{2}$ and $u_{1}$ in closed analytical form:
\begin{equation}
\label{real_omega}
{\Omega_{2}(r,z) \over \Omega_{0}}= \left({h\over r}\right)^{2}\left[
-{3 \over 4} + {1 \over 2}\left({d \ln{h} \over d \ln{r}}\right)+ {2
\over 15}\alpha^{2}\Lambda\left(1 - 6{z^{2} \over
h^{2}}\right)\right],
\end{equation}
\begin{equation}
\label{real_u}
u_{1}(r,z)= -\alpha\left({h^{2} \over r^{5/2}}\right)\left[- \Lambda
\left(1 - {z^{2}\over h^{2}}\right)- \Lambda\left({32 \alpha^{2} \over
15}\right) + 2\left({d \ln{h} \over d \ln{r}}\right)\right],
\end{equation}
where $\Lambda$ is a constant that depends on the parameter $\alpha$
and is given by
\begin{equation}
\label{Lambda_alpha}
\Lambda= \left.{11 \over 5}\right/ \left(1+ {64 \over
25}\alpha^{2}\right).
\end{equation}

Note that $\Omega_{2}(r,z)$
exhibits differential rotation with respect to the vertical
coordinate; a feature which was also observed in the numerical solution
by Kley \& Lin \markcite{KLN92}
(1992).This is because by including the viscous term of
$\eta_{0}(\partial u_{1}/ \partial z)$ in the radial momentum equation
we are no longer solving for $\Omega$ under an assumption of strict
radial hydrostatic equilibrium.

We also observe that in the limit of $\alpha \ll 1$, so that
$\alpha^{2}$ is a vanishingly small quantity,
eqs.~(\ref{real_omega})-(\ref{real_u}) reduce identically to the
solutions for $\Omega_{2}$ and $u_{1}$ of the equations of motion without
the coupling terms. This suggests
that  neglecting the viscous coupling terms is justified, so long as
$\alpha$ is not too large, i.e. $\alpha \le \epsilon$.
However, when $\alpha \sim 10^{-1}$ to 1, the effects
of including the ${\mathcal{O}}(\epsilon^{2})$ viscous terms becomes
readily apparent.

Note that $u_{1}(r,z)$, being quadratic in $z$, is an even
function with respect to reflections through the midplane $(z=0)$. It
is also clear, upon reinstating the appropriate scale factors, that
$v_{r}/v_{\phi k} \sim H^{2}/r^{2} \sim \epsilon^{2}$ in the outer
disk, where our use of eq. (\ref{lambda_true}) for the disk
surface is known to be valid. 
Since $(d \ln{h}/ d \ln{r})$ diverges at $r_{+}$, we observe
that $\lim_{r\rightarrow r_{+}} u_{1}(r,z)\rightarrow -\infty$,
i.e. we have not cured the well known divergence of the Shakura--Sunyaev
disk at the zero torque radius: as $r\rightarrow r_+$, $h\rightarrow 0$,
$\rho\rightarrow 0$ and, to preserve $\dot M=$const, $v_r\rightarrow \infty$.
However, for all radii $r>r_+$, $u$ is finite everywhere and on the
surface of the disk has the finite nonzero value given by
\begin{equation}
\label{no_div_u}
u_{1}(r,\pm h)= -\alpha \left({h^{2}\over
r^{5/2}}\right)\left[2\left({d \ln{h} \over d \ln{r}}\right)
-{352\alpha^2 \over 3 (25+64\alpha^2)} \right] <0.
\end{equation}

With the knowledge that $u_{1}$ remains finite at the surface of the
disk, we can now pursue, with confidence, a solution for the lowest
non-vanishing order of $v_{z}$, i.e. $v_{2}$, by means of
eq.~(\ref{simple_continuity}), the equation of continuity up through
second order in $\epsilon$. If we use eq.~(\ref{rho_0}) for $\rho_{0}$
and eq.~(\ref{real_u}) for $u_{1}$, then after the requisite
differentiation (with respect to $r$) and integration (with respect to
$z$), we can determine $v_{2}(r,z)$ up to an unknown function of the
disk radius. Yet since $v_{z}$ is an odd function of $z$ and therefore
$v_{2}=0$ in the equatorial plane for all $r$, this implies that the
unknown function of $r$ must be zero everywhere in the disk. In this
manner, we find the following unique solution for $v_{2}$:
\begin{equation}
\label{real_v}
v_{2}(r,z)= - \alpha \left({z \over r}\right)\left({h^{2} \over
r^{5/2}}\right)\left[-\Lambda\left(1 - {z^{2} \over h^{2}}\right)- {32
\alpha^{2} \Lambda \over 15}\left({d \ln{h} \over d \ln{r}}\right)+
2\left({d \ln{h} \over d \ln{r}}\right)^{2} \right]
\end{equation}
where $\Lambda$ is the same constant that appears in
eq.~(\ref{Lambda_alpha}). 

We immediately notice several
important features of the solution. First, as with $\Omega_{2}(r)$ and
$u_{1}(r,z)$, $\lim_{r \rightarrow r_{+}}v_{2}(r,z)\rightarrow
-\infty$ because of its dependence on $(d \ln{h}/ d
\ln{r})^{2}$. Secondly, we see that $\vert v_{z}/v_{r} \vert \sim H/r
\sim \epsilon$, as was expected for a standard thin disk in vertical
hydrostatic equilibrium. Finally, $v_{2}$, like $u_{1}$, is finite
along the surface of the disk for all $r$, i.e.
\begin{equation}
\label{no_div_v}
v_{2}(r,\pm h)= \mp \alpha \left({h^{3} \over
r^{7/2}}\right){d \ln{h}\over d \ln{r}}
\left[2\left({d \ln{h}\over d \ln{r}} \right)
-{352\alpha^2 \over 3 (25+64\alpha^2)} \right].
\end{equation}
We now check our solutions to see if they concur with what we expect
for a standard thin disk. First, it is possible to show that
eq.~(\ref{real_u}) satisfies eq.~(\ref{scaled_vert_angular_1}) for the
vertically-integrated mass flux, $\dot{m}$, even with the unexpected
dependence on $\alpha^{2}$. Second, a quick glance reveals that our
new solutions for $\Omega_{2}$, $u_{1}$, and $v_{2}$ have the
necessary parities with respect to reflection through the $(z=0)$
midplane, i.e. even, even, and odd, respectively. Third, if we replace
the appropriate dimensional units for each quantity, we see that
$(v_{\phi}- v_{\phi k})/v_{\phi k}$ and $v_{r}/v_{\phi k}$ are both of
${\mathcal{O}}(\epsilon^{2})$, while $v_{z}/v_{\phi k}$ is
${\mathcal{O}}(\epsilon^{3})$; all of which is entirely in complete
agreement with our assumption that the flow in the outer portions of
the disk is predominantly Keplerian.

Finally, since $h(r)$ is the semi-thickness of the disk, we must have
$v_{z}/v_{r}=\pm(dh/dr)$ for all points $(r,\pm h)$ on
the surface of the disk, as otherwise the geometrical disk surface
cannot be in a steady-state. Comparison of eqs.~(\ref{no_div_u}) \&
(\ref{no_div_v}) shows that this constraint
is indeed satisfied for the lowest non-vanishing orders of $v_{r}$ and
$v_{z}$:
\begin{equation}
\label{surface_check}
\pm{ v_{2}(r, \pm h) \over u_{1}(r,\pm h)}= \left({h \over
r} \right)\left({d \ln{h} \over d \ln{r}} \right)= {d h \over d r}.
\end{equation}
We stress that the new solutions for $\Omega_{2}$, $u_{1}$, and
$v_{2}$ fully comply with the above list of conditions for all
possible values of $\alpha$, provided that the disk remains
geometrically thin. The remaining properties of
eqs.~(\ref{real_omega})-(\ref{no_div_v}) for $\Omega_{2}$, $u_{1}$, and
$v_{2}$ will be discussed in detail in the following section.

\section{Detailed discussion of analytical results. {\label{Ch4}}}
\thispagestyle{myheadings}
\bigskip

In this section we examine the detailed
properties of $\Omega_{2}$ (eq.~[\ref{real_omega}]),
and in $\S$\ref{vec_field_1} those of
the remaining components of the vector field,
$u_{1}$, and $v_{2}$,
eqs.~(\ref{real_u})-(\ref{no_div_v}).  Since by assumption ii)
($\S\S$\ref{assumptions}, \ref{prelims})
the next order corrections
vanish everywhere in the disk, $\Omega_{3}=u_{2}=v_{3}=0$,
our equations are valid 
up to corrections of
${\mathcal{O}}(z^{4}/r^{4})$ for $\Omega$ and $v_{r}$, and
${\mathcal{O}}(z^{5}/r^{5})$ for $v_{z}$. In other words, the results
we describe are valid in the outer disk ($r>1.4r_+$) up to relative
corrections of $\epsilon^2$, e.g. for $H/R\approx0.1$ the expressions are
valid to 1\%.

In $\S$~\ref{omega_2_study},
we study the nature of the $z$ dependence of
$\Omega_{2}(r,z)$. In $\S$~\ref{singular_omega}
we also discuss the interpretation
of the apparent singularity in $\Omega_{2}$ at $r=r_{+}$. In
$\S$~\ref{vec_field_1}, we look at the velocity vector field in the
outer disk, paying special attention to the sign of
$v_{r}$ and $v_{z}$. We find that beyond a certain radius there is
significant mass outflow near the equatorial plane for a wide range of
$\alpha$'s. We
carefully analyze all of its important characteristics, including, in
$\S$~\ref{outflow_1}, the fraction of the total mass flow carried out
to infinity by this phenomenon.

\subsection{Analysis of the lowest-order correction to $\Omega$.
 {\label{omega_2_study}}}
\bigskip

We found that the angular velocity of
material in the disk, up to and including terms of
${\mathcal{O}}(\epsilon^{2})$ is given by
\begin{equation}
\label{total_omega}
\Omega=\Omega_{0}\left\{1+ \epsilon^{2}\left({h \over r }
\right)^{2}\left[-{3 \over 4} + {1 \over 2}\left({d \ln{h} \over d
\ln{r}}\right)+ {2 \over 15}\alpha^{2}\Lambda\left(1 - 6{z^{2} \over
h^{2}}\right)\right]\right\}.
\end{equation}
Eq.~(\ref{total_omega}) for $\Omega$ is valid up to
corrections
of ${\mathcal{O}}(z^{4}/r^{4})$ for a geometrically thin disk. 

As pointed out in Ch.~\ref{Ch3} 
$\Omega$ is clearly an even function of $z$ and is
Keplerian up through first order in $\epsilon$. In addition, since
${\lim_{r \rightarrow \infty}}(\dlnh) = 1$ and $\lim_{r \rightarrow
\infty}(h/r) = \lambda$ from 
eq.~(\ref{lambda_true}), we find that
\begin{equation}
\label{limit_omega}
\lim_{r \rightarrow \infty} \left({\Omega \over \Oo}-1 \right)=
\epsilon^{2} \lambda^{2}\left[-{1 \over 4} + {2 \over
15}\alpha^{2}\Lambda \left(1-6{z^{2} \over h^{2}} \right) \right].
\end{equation}
Note that this limit is negative for all $z$ because $0\le 2
\alpha^{2}\Lambda /15 \le 0.082$ over the range \linebreak $0\le
\alpha \le 1$. Also observe that in the case of $\alpha \ll 1$,
eq.~(\ref{limit_omega}) reduces to a negative constant
 $-\epsilon^{2} \lambda^{2}/4$.
Thus, in the limit of inviscid flow ($\alpha\rightarrow0$),
$\Omega$ is constant on cylinders (but subkeplerian for $r>>r_+$). Finally,
because $h^{2} (\dlnh) \rightarrow \infty$ as $r \rightarrow r_{+}$,
we know that $\Omega$, as given by eq.~(\ref{total_omega}), diverges
at $r=r_{+}$.  It is on these last two properties that we now focus
our attention.

\subsection{Differential rotation of the angular velocity with respect to
 $z$. {\label{differ_z}}}
\bigskip

In this subsection, we discuss the $z$ dependence of
$\Omega$. We define the quantity $\Delta
\Omega=[\Omega(r, \pm H)- \Omega(r,0)]$ to be the difference between
the angular velocity at the surface of the disk and the angular
velocity in the equatorial plane. From
eq.~(\ref{total_omega}), we derive the following expression for the
fractional difference:
\begin{equation}
\label{frac_omega}
(\Delta \Omega/ \Oo) = -{\epsilon^{2}\over \Oo}\left[\Omega_{2}(r,\pm
h)-\Omega_{2}(r,0) \right]=-\epsilon^{2}\left({h \over
r}\right)^{2}\left({4 \over 5} \alpha^{2} \Lambda\right).
\end{equation}
Clearly, this fraction is negative for all radii. In Fig.~1, we plot
$\log{\vert \Delta \Omega /\Oo \vert}$ in the disk with $\epsilon =
0.01$ and $\dot{m}=1$, for several values of the viscosity parameter:
$\alpha = 0.01$, $0.1$, and $1.0$. We observe that in
the limit of $r \gg r_{+}$, since $\lim_{r \rightarrow \infty} (h/r) =
\lambda$ {\mbox{(see eq. (\ref{dimensionless_h}))}}, the fractional
difference in $\Omega$ tends to a negative constant given by
{\mbox{$\lim_{r \rightarrow \infty} (\Delta \Omega/ \Oo) \rightarrow
-\epsilon^{2}(4/5)\alpha^{2}\Lambda \lambda^{2}$}} (dashed lines).

The amount by which the surfaces of constant $\Omega$ deviate from
upright cylinders is proportional to $\epsilon^{2}$ and therefore very
small for a thin disk. The nearly constant behavior for $(\Delta
\Omega/ \Oo)$ for $r \gg r_{+}$ is also interesting in its own right
since it implies that the total departure of the 
surfaces of constant $\Omega$ from the vertical does not
significantly decrease (or increase) as one goes further out in the
disk. Examination of eq.~(\ref{frac_omega}), also shows us that formally
differential rotation vanishes as $r\rightarrow r_{+}$,
since $h\rightarrow 0$ at $r=r_{+}$; however, as $\Omega= \Oo +
\epsilon^{2}\Omega_{2}$ diverges to $+\infty$ at $r=r_{+}$, this
result of $\Delta \Omega \rightarrow 0$ really has no physical
meaning.

\subsection{The singularity in $\Omega(r,z)$ in the inner disk.
{\label{singular_omega}}}
\bigskip

In Fig.~2(a), we plot the second order correction to $\Omega$
(i.e. $\epsilon^{2}\Omega_{2}$) as a function of $r$ in the $(z=0)$
midplane for three different values of $\alpha$. In Fig.~2(b), we
plot $H= \epsilon h(r)$ for the same values $\alpha$ used in
Fig.~2(a). For each choice of $\alpha$, observe the singularity in
$\Omega_{2}$ at $r_{+}$.
 Clearly, as the presumed $\Ord2$
correction provided by $\Omega_{2}$ grows without bound near
$r=r_{+}$, the assumptions underlying
our perturbative expansion are invalid there.

The reason for the divergence, at  $r=r_{+}$,
of $\Omega_2$ , $v_{r}$ and $v_{z}$ can be traced back
directly to the lowest order form of the vertically-integrated angular
momentum equation (\ref{scaled_angular_integrated}):
\begin{equation}
\label{reprint_scaled_vert_angular}
\dot{m} (j_{0}-j_{+}) = -r^{3}\left[\int_{-h}^{+h} \eta_{0}dz\right]
{d \Omega_{0}\over d r}.
\end{equation}
The left side of eq.~(\ref{reprint_scaled_vert_angular})
is proportional to $(j_{0}-j_{+})=j_{0}\left(1-\sqrt{r_{+}/r}\right)$
and as such must vanish at $r=r_{+}$. Since we are still assuming that
$\Omega \approx \Oo$ in the neighborhood of $r_{+}$, if the right side
is to also vanish there, then it is necessary for the
height-integrated viscosity, $\int_{-h}^{+h} \eta_{0} dz$, to go to
zero at $r=r_{+}$.  However, in the \linebreak $\alpha$-disk
prescription $\intinfh \eta_{0}dz$ is proportional to $
(h/r)^{6}$, so this leads to the conclusion that $h\rightarrow 0$ as
$r\rightarrow r_{+}$ (Shakura \& Sunyaev 1973). 
This functional form of $h(r)$ causes a
singularity in its derivative at $r_{+}$,
i.e.~{\mbox{$\lim_{r\rightarrow r_{+}}(d h/ dr)\rightarrow +\infty$}},
and hence to a  ``cusp'' in the disk surface.

The fault for all of this lies in the assumption that one can
continue the perturbative expansion into the transition region, near
the zero-torque point, $r_{max}$, where $\Omega$ reaches a
maximum. While it is true that eq.~(\ref{raw_vert_angular_final})
holds for all radii, including $r=r_{+}$, the same cannot be said for
its lowest order expansion in $\epsilon$,
eq.~(\ref{reprint_scaled_vert_angular}), which was used in the
derivation of $h(r)$ and hence all other physical quantities. Since
$\Omega$ is maximal at $r=r_{max}$, $\partial \Omega / \partial r
\rightarrow 0$ there, and eq.~(\ref{raw_vert_angular_final}) is easily
satisfied without requiring that $\intinf \eta_{0} dz$ vanish anywhere
in the disk. However, 
eq.~(\ref{reprint_scaled_vert_angular}), presumes that $\Omega$ and
$j$ were approximately equal to their corresponding 
Keplerian values of $\Oo$ and $j_{0}$. This then forced the
vertically-integrated viscosity (and hence $h$) to be zero at the
radius, $r_{+}$; ultimately leading to the calculated singularities in
$\Omega$, $v_{r}$ and $v_{z}$.

In the final analysis, we arrive at the conclusion that our
perturbative expansion in $\epsilon$ is not valid in the inner part of
the disk, near $r=r_{+}$, if we use the simple analytical expression
for $h(r)$ which appears in eq.~(\ref{dimensionless_h}). However,
as they are written in terms of the disk surface, $h(r)$, our
solutions for $\Omega_{2}$, $u_{1}$, and $v_{2}$ in
eqs.~(\ref{real_omega})-(\ref{real_v}) are more general than they
first appear. Numerical work (Kita \& Klu\'zniak 1997) shows that
the radius of convergence for our expansions can be extended well into
the boundary layer ($r_+\gtrsim r$) and all singularities completely disappear
if only a more realistic (numerical) solution for $h$ and
$(d \ln{h}/ d \ln{r})$ is used. In any
event, our solutions for the vertical structure are certainly valid in
the outer regions of the disk for $r \gtrsim 2 r_{+}$, where we are
fully justified in assuming $\Omega \approx \Omega_{k}$, and hence in
using eq.~(\ref{reprint_scaled_vert_angular}) to determine $h$.

\section{The velocity vector field. {\label{vec_field_1}}}
\bigskip

In this section we will examine the behavior of the horizontal and
vertical components of the fluid velocity in the accretion disk.
In particular, we will concern ourselves with the
sign of $v_{r}$ and $v_{z}$, so that we can determine the direction
of the accretion flow. Note that we adopt here the convention that
accompanied eq.~(\ref{basic_mdot}), i.e. $\dot{M} > 0$ for
accretion. Thus, if $v_{r}<0$ the radial component of the flow is
directed towards the central star.
Likewise, $v_{z}<0$ for $z>0$ signifies flow towards
the $z=0$ midplane. 
Eqs.~(\ref{real_u})-(\ref{no_div_v}) for $u_{1}$ and $v_{2}$ directly
give the unscaled velocity components as functions of $r$ and $z$
to lowest order in $\epsilon$, \linebreak
i.e. $v_{r}(r,z)=\epsilon^{2}u_{1}\Omega_{k*}\tilde R$ and
$v_{z}(r,z)=\epsilon^{3}v_{2}\Omega_{k*}\tilde R$.

\subsection{The sign of $v_{r}$ in the outer disk: outflow vs. inflow.
 {\label{u_out}}}
\bigskip 

It is a result of our analysis, that at the surface of the accretion
disk considered here, the fluid always flows in the general direction
of the central object. Indeed, since $2>(32/15)\Lambda \alpha^{2}$ for
all possible $\alpha$'s
in the range $0\le \alpha \le 1$, then $v_{r}<0$ for all points on the
disk surface because the term in eq.~(\ref{real_u})
proportional to $(1-z^{2}/h^{2})$ vanishes at $z=\pm h$.
However, near the midplane beyond a certain radius, 
there may or may not be outflow depending on the
chosen value of the parameter of viscosity, $\alpha$.

The radial velocity in the equatorial plane is:
\begin{equation}
\label{vr_eq}
v_{r}(r,0)=-\alpha \epsilon^{2}\Omega_{k*}\tilde R\left({h^{2} \over
r^{5/2}}\right)\left[2\Dlnh - \Lambda \left(1+{32 \over 15}
\alpha^{2}\right)\right],
\end{equation}
where $(\dlnh)$ can be obtained from eq.~(\ref{lambda_true}) and
$\Omega_{k*}\tilde R$ provides the physical units. For small radii,
$r\sim r_{+}$, evaluation of {\mbox{eq.~(\ref{vr_eq})}} shows that
the term involving $(\dlnh)$ dominates the bracketed expression and
thus $v_{r}(r,0)<0$, i.e. we have inflow. However, for large radii,
$r\gg r_{+}$, we see that since $\lim_{r \rightarrow \infty}(\dlnh) =
1$, the sign of $v_{r}(r,0)$ depends ultimately on whether or not $2
\ge \Lambda(1+32\alpha^{2}/15)$.

If, for example, $\alpha = 0$, then $\Lambda$ reduces to $11/5$
 and the bracketed portion of {\mbox{eq.~(\ref{vr_eq})}} is negative, so
that $v_{r}(r,0)>0$ for $r\gg r_{+}$; indicating the existence of
outflow in the equatorial plane far from the central
accretor{\footnote{To our knowledge, equatorial outflow in accretion
disks was first discovered by Urpin \markcite{URP84} (1984).}}.
On the other extreme, if
$\alpha=1$ then $\Lambda \approx 0.618$ and $\Lambda
[1+32\alpha^{2}/15]\approx 1.94 < 2$ and so $v_{r}(r,0)<0$ for $r \gg
r_{+}$, signaling that the equatorial flow is directed inwards towards
the central star for all radii. A more
precise analysis of {\mbox{eq.~(\ref{vr_eq})}} allows us to determine
the critical value $\alpha_{cr}=\sqrt{15/32}\sim 0.685$
above which there is no backflow.
This is a rather substantial value more than ten times as
large as the critical value estimated by Kley \& Lin {\markcite{KLN92}
(1992).

We now analyze quantitatively where the direction of the flow changes
sign when $\alpha < \alpha_{cr}$. To this end we denote the stagnation
radius, $r_{stag}$, as the radius for which $v_{r}=0$ in the
equatorial plane.  Eq.~(\ref{vr_eq}) with $\alpha <
\alpha_{cr}$ allows us to calculate this stagnation radius as a
function of $\alpha$:
\begin{equation}
\label{r_stagnation}
{r_{stag}(\alpha) \over r_{+}}={\left[ 1+6\left(\Lambda(1+{32 \over
15}\alpha^{2})-2\right)\right]^{2} \over \left[6\left( \Lambda(1+{32
\over 15}\alpha^{2})-2\right)\right]^{2}}.
\end{equation}
Observe that as $\alpha \rightarrow \alpha_{cr}=\sqrt{(15/32)}$,
$\Lambda \rightarrow 1$ by eq.~(\ref{Lambda_alpha}), and
$r_{stag}\rightarrow \infty$, as expected.
This behavior is clearly visible in Fig 3, where we
have plotted $\log{(r_{stag}/r_{+})}$ vs. $\alpha$.
In the opposite limit $\alpha\rightarrow0$,
$r_{stag}/r_+\rightarrow 121/36\approx3.36$.
For $\alpha \le
10^{-1}$, we see that $r_{stag}\approx 3.5 r_{+}$, this suggests that
there is mass outflow in the midplane throughout most of the disk
for realistic values of $\alpha$.

Of course, for $\alpha < \alpha_{cr}$, the region of
outflow is not restricted to the midplane $(z=0)$ . Indeed,
inspection of eq.~(\ref{real_u}) shows that $v_{r}(r,z)\ge 0$ for
a range of $z$ above and below the equator for $r\ge r_{stag}$. We,
therefore, define the ``vertical flow surface,'' $z_{vert}(r,\alpha)$,
as the surface on which $v_{r}=0$, implying that the flow is moving
vertically there and hence the name. Clearly, $z_{vert}$ is
well-defined for $r \ge r_{stag}$ and intersects the equatorial plane
at $r=r_{stag}$.  Thus we can expect there to be outflow in the disk
for all points contained in the domain of $r > r_{stag}(\alpha)$ and
$z <z_{vert}(r,\alpha)$, whenever $\alpha \le \alpha_{cr}$. For
$\alpha \ge \alpha_{cr}$ the surface of vertical flow
disappears entirely.

Solving for $z_{vert}$ via eq.~(\ref{real_u}) for $u_{1}$, we obtain
\begin{equation}
\label{z_vertical}
\left({z_{vert}(r,\alpha) \over h(r)}\right)= \sqrt{\left(1+{32 \over
15}\alpha^{2}\right)-{2 \over \Lambda}\Dlnh}.
\end{equation}
In Fig.~4, we show $\log(Z_{vert}/r_{+})$ (solid curves) and
$\log(H/r_{+})$ (dashed curves) for two different values{\footnote{$H$
is also affected by the choice of $\alpha$, since $h \propto
(\dot{m}/\alpha)^{1/6}$.}} of the
parameter $\alpha$: $\alpha =0.1$ \& $\alpha=0.6$, where
$Z_{vert}=\epsilon z_{vert}$ and $H=\epsilon h$. Observe that in
Fig.~4, since $\alpha=0.6$ is closer to $\alpha_{cr}\sim 0.685$, the
outflow region, bounded from above by $z_{vert}(r,\alpha)$, is
beginning to be strongly suppressed and does not even start in the
equatorial plane until $r \gtrsim 60 r_{+}$. However, for
$\alpha=0.1$ in Fig.~4, the region of backflow is quite
extensive and occupies nearly $30\%$ of the disk for $r\ge 10 r_{+}$.

\subsection{The sign of $v_{z}$ and its impact on mass outflow.
 {\label{v_out}}}
\bigskip

We turn now our attention to the vertical velocity component,
$v_{z}$. A study of eq.~(\ref{real_v}) reveals that $\pm v_{z}<0$
for all points on the disk surface at $z = \pm h(r)$, respectively,
signifying that the flow there is
directed towards the equatorial plane. In addition, eq.~(\ref{real_v})
leads us to the conclusion that for $\alpha < \alpha_{cr}$, besides
being zero for all radii in the $(z=0)$ midplane because of its odd
parity, $v_{z}$ also vanishes on a new and different ``horizontal
flow surface'', $z_{hor}(r,\alpha)$, on which the flow
is directed horizontally.

We determine the following
functional form for this surface:
\begin{equation}
\label{z_horiz}
\left({z_{hor}(r,\alpha) \over h(r)}\right)= \sqrt{\left(1+{32 \alpha^{2}\over
15}\cdot\Ddlnh\right)-{2 \over \Lambda}{\Dlnh}^{2}}.
\end{equation} 
Comparison of this result with eq.~(\ref{z_vertical}) reveals some
definite similarities between the two flow surfaces, including the
property that $z_{hor}$ vanishes entirely in the limit $\alpha
\rightarrow \alpha_{cr}$. We also observe that $z_{hor}$, the bounding
surface across which $v_{z}$ changes sign, is contained entirely
within the vertical flow surface, $z_{vert}$, at which
$v_{r}=0$. Furthermore, as $r\rightarrow \infty$, $z_{hor}$
asymptotically approaches $z_{vert}$ from below for all values of
$\alpha$. Thus for $\alpha < \alpha_{cr}$, there is a nested volume
contained within the domain of backflow ($v_{r}>0$), within which
the flow is directed {\underline {away}} from the
midplane,i.e.  $v_{z}/z >0$.

Figs.~5(a)-(b), illustrate this phenomenon for $\alpha=0.1$ and
$\alpha=0.6$ respectively, where we have zoomed in to that fraction of
the disk involved with outflow. Observe that at smaller radii, in both
cases region $A$, the area within which $v_{z}/z>0$, is much
smaller than the corresponding region $B$ within which
$v_{r}>0$; though their boundaries do converge to another at infinity,
since $\lim_{r \rightarrow\infty} (d \ln{h} / d \ln{r}) \rightarrow 1$.

Eqs.~(\ref{real_u})-(\ref{real_v}), in conjunction with
eqs.~(\ref{z_vertical})-(\ref{z_horiz}), also allow us to draw some
rather interesting conclusions regarding the flow geometry in the
outer disk. First, any material that enters region $B$ {\underline
{must}} cross over into region $A$ at some time in the future. To see
this, note that $v_{z}/z$ is still negative for all points between
the vertical flow surface and the horizontal flow surface
($|z_{hor}|<|z|<|z_{vert}|)$ and thus any fluid
element in region $B$ (where $v_{r}>0$), given enough time, must
eventually cross the inner horizontal flow surface, $z_{hor}$,
and into region $A$. It is clear that some of the material in region $C$ in
Figs.~5(a)-(b) must {\underline {not}} enter region $B$, instead a
sizeable fraction of the flow that exists above $z_{vert}$ must
continue on ahead of the stagnation radius, into region $D$ and onto
the central star. This is, of course, the accreted flow that comprises
$\dot{M}$.

To better visualize our previous remark, we call the reader's
attention to Figs.~6(a)-(b), where we plot the direction of the
velocity vector field, i.e. the unit vectors formed from the
components $v_{r}$ and $v_{z}$, in the outer disk for two different
values of viscosity, both with $\alpha < \alpha_{cr}$.
As a reference, we have also overlaid the vertical
flow surface, $z_{vert}$. In Fig.~6(c), we show the directional flow
pattern for the case of $\alpha > \alpha_{cr}$ (here $\alpha =1$), and
we now find that only inflow persists throughout the entire disk.

Finally,  note that there are no cells of meridional circulation in the disk.
In the region $B$ flow is always directed towards the surface which
bounds region $A$, while in the region $A$ flow is always
directed away from the equator. The circulating
pattern of flow found by Kley \& Lin (1992) close to the edges of
their computational domain must be an artifact of the boundary conditions
they imposed.
In fact, calculation of the stream lines shows that material which
enters region $A$ can {\underline {never}} intersect the $v_{z}=0$
horizontal flow surface again, because $v_{r}>0$ and $v_{z}
\rightarrow 0$ there. Coupled with our original finding that any fluid
which enters region $B$ must also pass eventually into region $A$,
this implies
that material once in the backflow region $B$ can never return to
the inflow portion $C$ of the accretion disk. Therefore, if
$\alpha < \alpha_{cr}$, the backflow must continue
all the way out to ``infinity,'' where the disk terminates.
All of this can be easily verified by looking at Figs.~6(a)-(b).

\subsection{The mass fraction of outflow relative to inflow.
{\label{outflow_1}}}
\bigskip

The last question that we address relates to the backflow.
What is the mass {\underline{outflow}}
rate in comparison to the $\dot{M}$ accreted by the star? It is clear
that the answer to this question must depend on the parameter
$\alpha$, since we know that there is no backflow for
$\alpha\ge \alpha_{cr}\sim 0.685$. For this reason we now define
$\Gamma(r,\alpha)$ to be the ratio of the mass rate flowing outwards through
a cylinder of radius $r>r_{stag}$ to the net mass accretion rate,
$\dot{M}$.

To evaluate the functional form of $\Gamma
(r,\alpha)$, we simply integrate the radial mass flux, $\rho v_{r}$,
over the surface of a cylinder with height $2z_{vert}$ centered on the
midplane. 
\begin{equation}
\label{Gamma_def}
\Gamma(r,\alpha)=\left|{4 \pi
r\int_{0}^{z_{vert}(r,\alpha)}\rho_{0}u_{1}dz \over 4 \pi
r\int_{0}^{h(r)}\rho_{0}u_{1}dz }\right|={1 \over
\dot{m}}\left|{2r\int_{0}^{z_{vert}(r,\alpha)}\rho_{0}u_{1}dz}\right|,
\end{equation}
where $\dot{m}$ is the scaled mass accretion rate defined in
eq.~(\ref{scaled_vert_angular_1}).

Using eq.~(\ref{z_vertical}) for $z_{vert}(r,\alpha)$,
eq.~(\ref{rho_0}) for $\rho_{0}$, and eq.~(\ref{real_u}) for
$u_{1}(r,z)$ we find that the mass outflow fraction is
\begin{equation}
\label{gamma_r_alpha}
\Gamma(r,\alpha)=\left({2\alpha \Lambda \over
5\sqrt{5} \dot{m}}\right)\left({h \over r}
\right)^{6}\left[G_{1}(\gamma^{*})+\left((\gamma^{*})^{2}-1\right)G_{2}(\gamma^{*})\right]
\end{equation}
where $\gamma^{*}=\gamma^{*}(r,\alpha)=z_{vert}/h$ and $G_{1}$ and
$G_{2}$ are rather complicated functions of $\gamma^{*}$ given by:
\begin{eqnarray}
\label{g1}
\nonumber G_{1}\gm = \int_{0}^{\gamma^{*}} \left(1-{z^{2} \over
h^{2}}\right)^{5/2}{dz \over h}= {5 \over 16}\sin^{-1}\gm + {5 \over
16}\gamma^{*}\left(\sqrt{1-{\gm}^{2}}\right)\hspace*{2.0cm}\\ + {5
\over 24}\gamma^{*}\left(\sqrt{1-{\gm}^{2}}\right)^{3}+ {\gamma^{*}
\over 6}\left(\sqrt{1-{\gm}^{2}}\right)^{5}, \hspace*{0.5cm}
\end{eqnarray}
\begin{eqnarray}
\label{g2}
\nonumber G_{2}\gm = \int_{0}^{\gamma^{*}} \left(1-{z^{2} \over
h^{2}}\right)^{3/2}{dz \over h}= {3 \over 8}\sin^{-1}\gm + {3 \over
8}\gamma^{*}\left(\sqrt{1-{\gm}^{2}}\right)\hspace*{2.5cm}\\ + {1
\over 4}\gamma^{*}\left(\sqrt{1-{\gm}^{2}}\right)^{3}.\hspace*{4.0cm}
\end{eqnarray}
The limit $\Gamma_{\infty}(\alpha)=\lim_{r\rightarrow \infty}
\Gamma(r,\alpha)$
is of particular interest, as it
represents the ratio of the {\underline {net}} outflow to net inflow
in the disk.

Analysis of eqs.~(\ref{gamma_r_alpha})-(\ref{g2}) 
provides us with the following form for $\Gamma_{\infty}$:
\begin{equation}
\Gamma_{\infty}(\alpha)={32 \Lambda \over
\pi}\left[G_{1}(\gamma^{*}_{\infty})+\left(\left(\gamma^{*}_{\infty}\right)^{2}-1
\right)G_{2}(\gamma^{*}_{\infty}) \right],
\end{equation}
where the quantity $\gamma^{*}_{\infty}$ is well defined only for
$\alpha < \alpha_{cr}$ and given by
\begin{equation}
\gamma^{*}_{\infty}= \lim_{r\rightarrow \infty}\gamma_{*}(r,\alpha)=
\sqrt{1+{32 \over 15}\alpha^{2}-{2 \over \Lambda}}.
\end{equation}

The resultant dependence on $\alpha$ for the mass outflow fraction as
$r\rightarrow \infty$ is plotted in Fig.~7. We see that 
$\Gamma_{\infty}(\alpha)$ tends smoothly to zero as
$\alpha \rightarrow \alpha_{cr}$. However, for a wide range of
$\alpha$, the fraction of the mass flow that is contained in the
outflow region, near the midplane is quite significant:
$\Gamma_{\infty}(\alpha)\sim 0.4$ for $0\le \alpha \lesssim0.1$ and
$0.1 <\Gamma_{\infty}(\alpha)\lesssim 0.35$ for $0.1<\alpha
<0.5$. Thus, for $\alpha\le0.1$,
the total mass rate, $\dot{M}_{tot}=\dot{M}_{out}+\dot{M}$, being
fed into the disk at the outer edge is $\approx 1.4 \dot{M}$.

All in all, we are led to the startling conclusion (Urpin 1984)
that, for small
values of $\alpha$, there is backflow in the disk transporting fluid
outwards to the outer boundary of the accretion disk. This should have
serious repercussions for
mass transfer at the outer edge of the disk.

\vfill\eject
~\par\noindent
{\bf 6. Final remarks.}
\smallskip\par\noindent
By performing a systematic expansion (pioneered by Regev 1983)
of the disk equations of motion we were able to find a closed solution
for the velocity field and the disk structure, valid everywhere outside,
say, 1.1 times the zero torque radius. We showed how, for all but very large
values of viscosity, the accretion flow turns around and feeds a backflow
(first discovered by Urpin in 1984) in the equatorial plane of the disk.
This backflow has now also been seen in a number of numerical simulations
and must be considered a general feature of accretion in a geometrically
thin disk, and possibly also in quasi-spherical flows.

We note, that if the flow discussed here
were advective, some of the gravitational
energy released by the flow close to the central gravitating body would
have been carried by the flow to larger radii before it is radiated. An urgent
topic of investigation should be whether
solutions with backflows, similar to the one presented here, may be
present in an advection dominated flow.
Should such solutions exist, conclusions (e.g. Narayan 1996)
that the apparent deficit of emission
in the inner region of accretion disks of some X-ray ``novae''
 necessarily implies
the presence of a space-time horizon would have to be treated with
caution.

\smallskip\par\noindent

\vfill\eject
\par\noindent REFERENCES

\par\noindent
Frank, J., King, A.R. \& Raine, D.J. 1992, Accretion Power in Astrophysics,

 Cambridge University Press.

\par\noindent
H\={o}shi, R. 1977, Prog. Theor. Phys. 58, 1191.

\par\noindent
Igumenshchev, I.V., Chen, X. \& Abramowicz, M.A. 1995, MNRAS 278, 236--250.

\par\noindent
Kita, D. B. 1995, Ph. D. Thesis, University of Wisconsin-Madison.

\par\noindent
Kita, D. B. \& Klu\'zniak W. 1997, in preparation.

\par\noindent
Klu\'zniak W. 1987, Ph. D. Thesis, Stanford University.

\par\noindent
Kley, W. \& Lin, D.N.C. 1992, ApJ 397, 600--612.

\par\noindent
Landau, L.D. \& Lifshitz, E.M. 1959, Fluid Mechanics, Pergamon Press.

\par\noindent
Monin, A.S. \& Yaglom, A.M. 1965, Statistical Fluid Mechanics: 1,
 Nauka: Moscow

 [MIT Press, 1971].

\par\noindent
Narayan, R. 1996, ApJ 462, 136--141.

\par\noindent
Narayan, R. \& Yi 1995, I. ApJ 444, 231--243.

\par\noindent
Paczy\'nski 1991, ApJ 370, 597--603.

\par\noindent
Prendergast, K.H. \& Burbidge, G.R. 1968, ApJ 151, L83--L88.

\par\noindent
Regev, O. 1983, AA 126, 146--151.

\par\noindent
R\'o\.zyczka, M., Bodenheimer, P. \& Bell, K.R. 1994, ApJ 423, 736--747.

\par\noindent
Shakura, N.I. \& Sunyaev, R.A. 1973, AA 24, 337--355.

\par\noindent
Tassoul, J.L. 1978, Theory of Rotating Stars, Princeton University Press.

\par\noindent
Urpin, V.A. 1984, Astron. Zh. 61, 84--90 [Sov. Astron. 28, 50--53].

\par\noindent
Urpin, V.A. 1984b, Astrophys. Sp. Sci. 90, 79.


\thispagestyle{myheadings}
\begin{figure}[h!]
\figurenum{1} \epsfysize=5.1875in {\vspace*{1.0in}
\epsffile{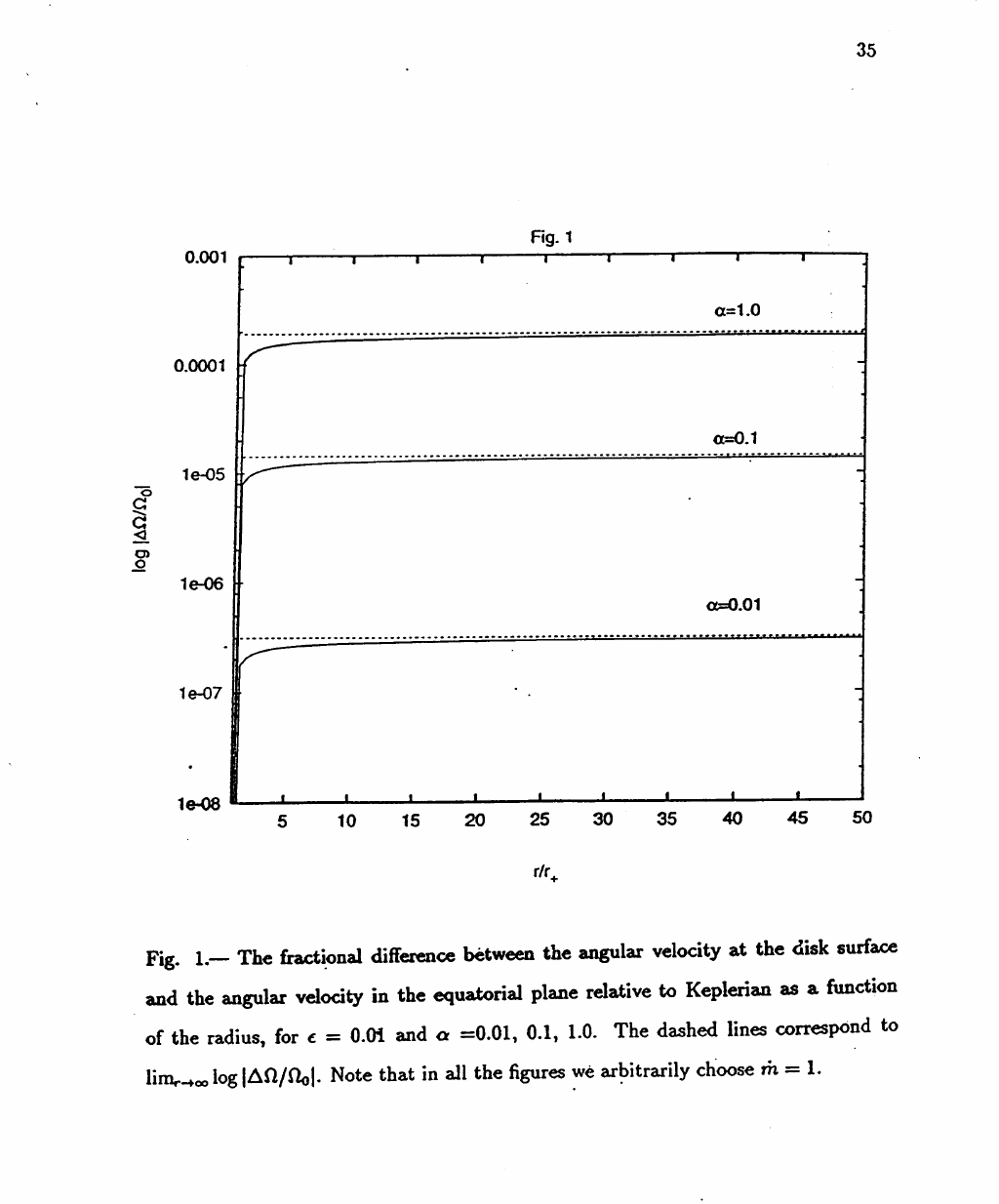}}
\caption{The fractional difference between the angular velocity at the
disk surface and the angular velocity in the equatorial plane relative
to Keplerian as a function of the radius, for $\epsilon=0.01$ and
$\alpha=$0.01, 0.1, 1.0. The dashed lines correspond to $\lim_{r
\rightarrow \infty} \log{\vert \Delta \Omega / \Omega_{0}\vert}$. Note
that in all the figures we arbitrarily choose $\dot{m}=1$.}
\end{figure}
\thispagestyle{myheadings}
\begin{figure}[h!]
\figurenum{2(a)} \epsfysize=5.1875in {\vspace*{1.0in}
\epsffile{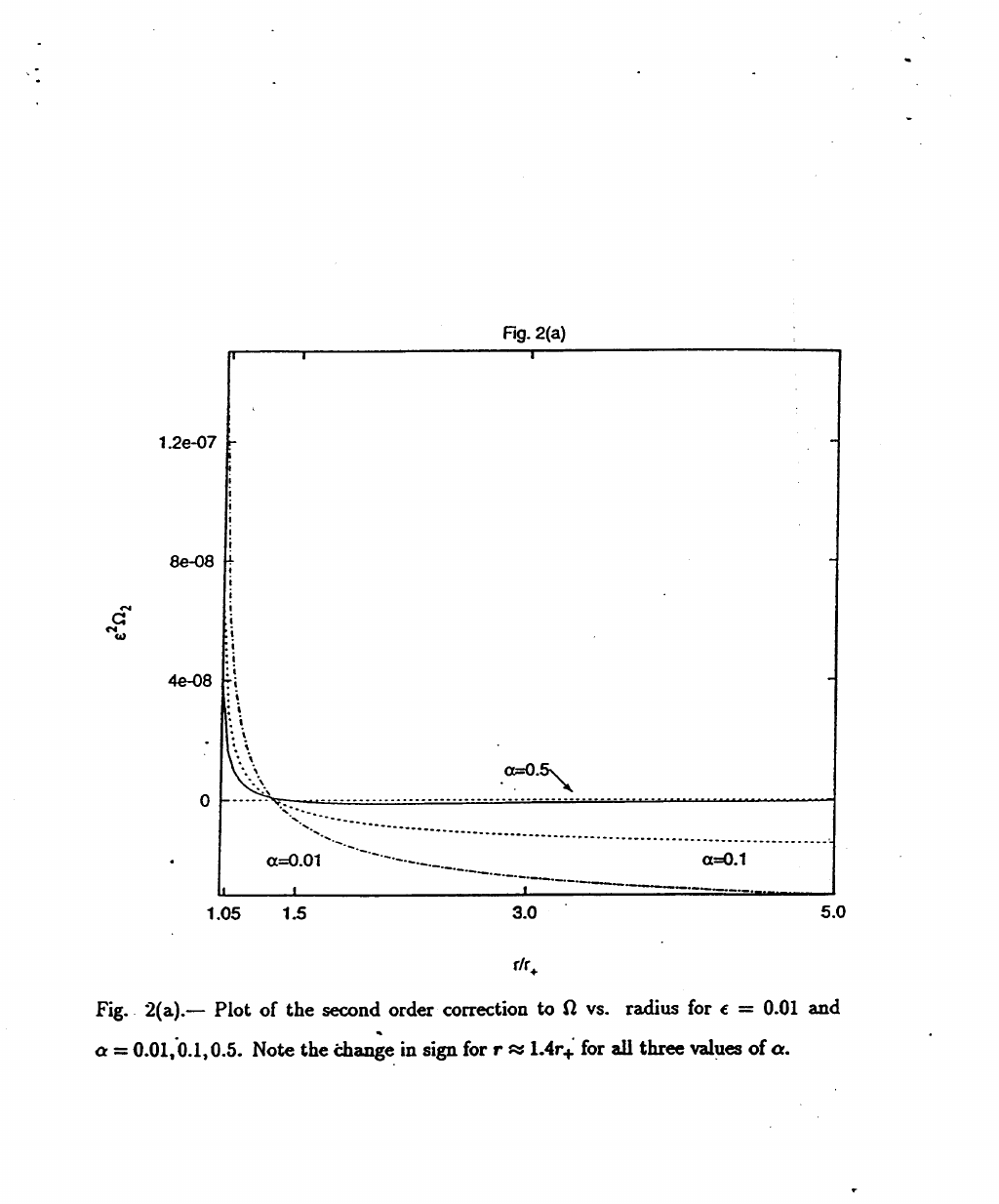}}
\caption{Plot of the second order correction to $\Omega$ vs. radius
for $\epsilon=0.01$ and $\alpha=0.01, 0.1, 0.5$. Note the change in
sign for $r \approx 1.4 r_{+}$ for all three values of $\alpha$.}
\end{figure}
\thispagestyle{myheadings}
\begin{figure}[h!]
\figurenum{2(b)} \epsfysize=5.1875in {\vspace*{1.0in}
\epsffile{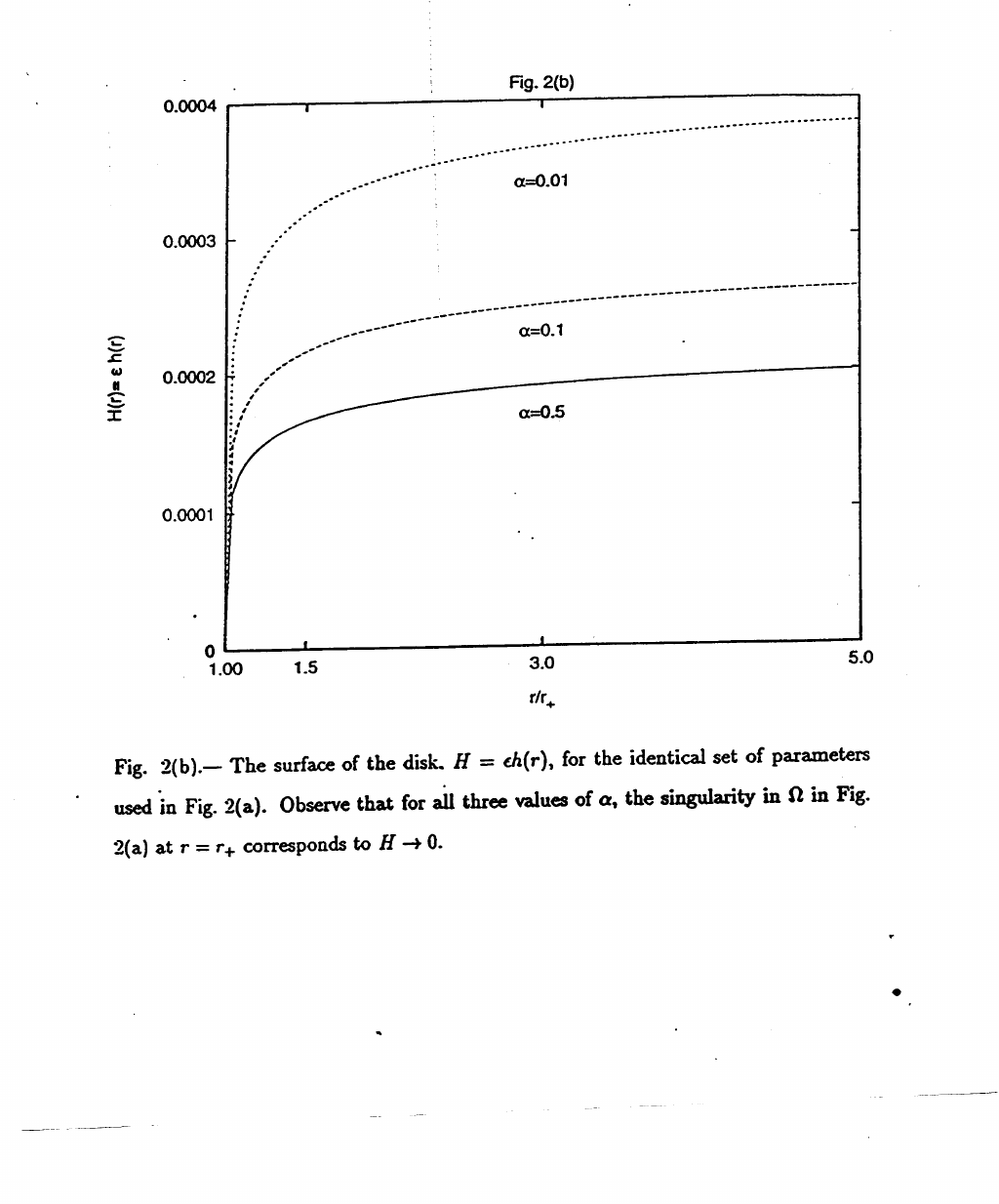}}
\caption{The surface of the disk, $H=\epsilon h(r)$, for the identical
set of parameters used in Fig.~2(a). Observe that for all three values
of $\alpha$, the singularity in $\Omega$ in Fig. 2(a) at $r=r_{+}$
corresponds to $H \rightarrow 0$.}
\end{figure}

\thispagestyle{myheadings}
\begin{figure}[h!]
\figurenum{3} \epsfysize=5.1875in {\vspace*{1.0in}
\epsffile{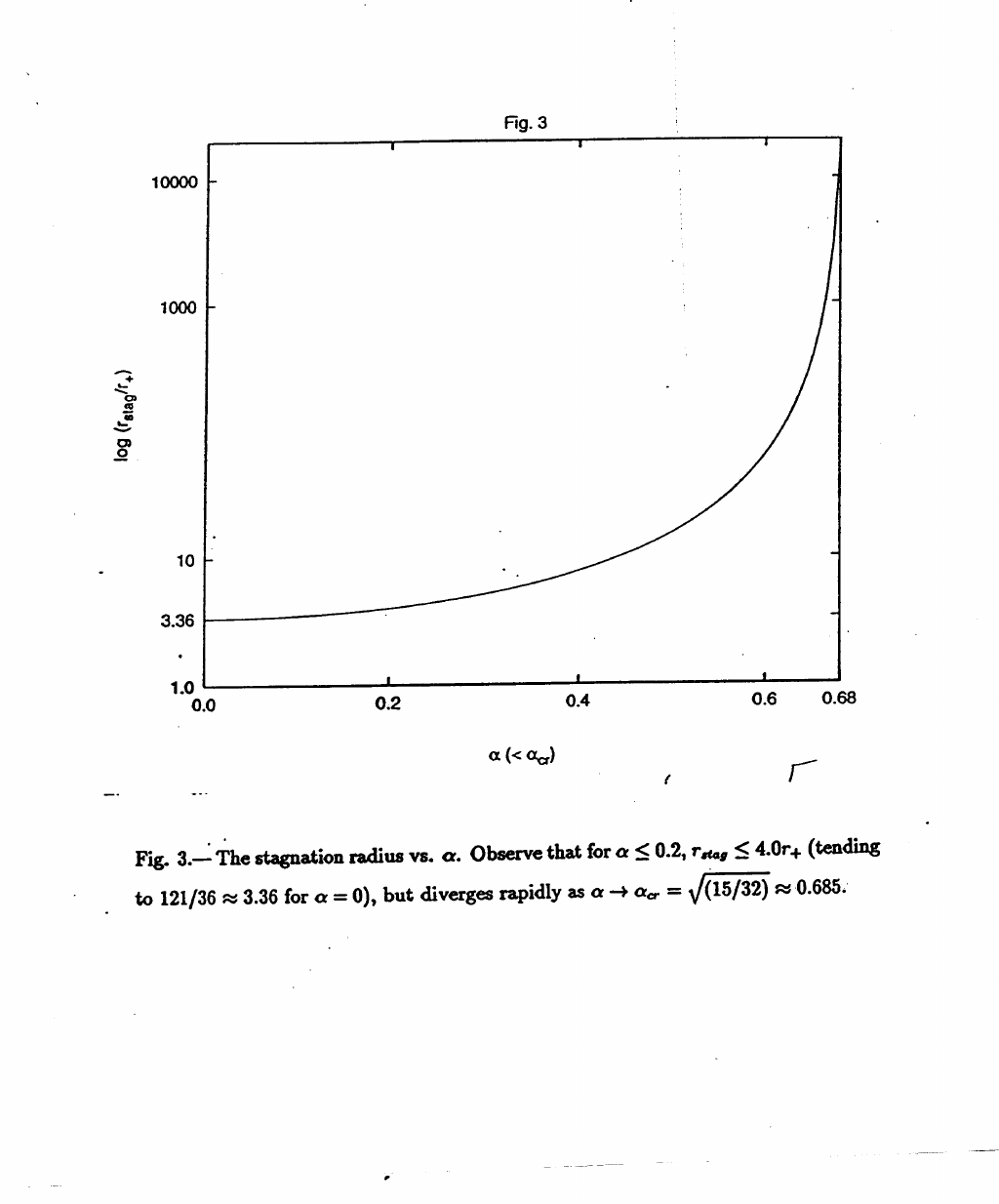}}
\caption{The stagnation radius vs. $\alpha$. Observe that for 
$\alpha\le0.2$,  $r_{stag}\le4.0 r_{+}$
(tending to $121/36\approx3.36$ for $\alpha=0$),
but diverges rapidly as $\alpha \rightarrow \alpha_{cr}=\sqrt{(15/32)}
 \approx0.685$.}
\end{figure}
\thispagestyle{myheadings}
\begin{figure}[h!]
\figurenum{4} \epsfysize=5.1875in {\vspace*{1.0in}
\epsffile{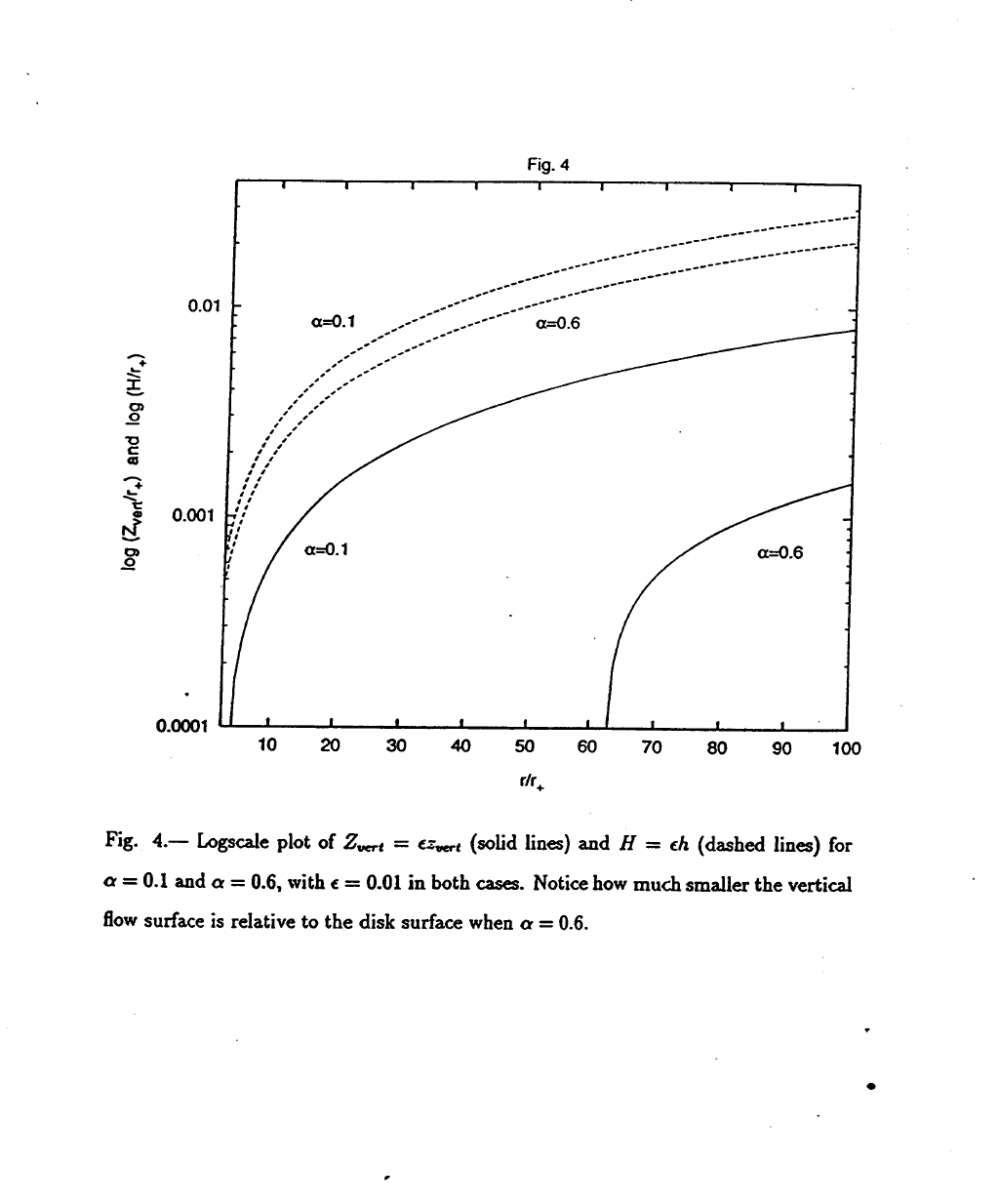}}
\caption{Logscale plot of $Z_{vert}=\epsilon z_{vert}$ (solid lines)
and $H=\epsilon h$ (dashed lines) for $\alpha=0.1$ and $\alpha=0.6$,
with $\epsilon=0.01$ in both cases. Notice how much smaller the
vertical flow surface is relative to the disk surface when
$\alpha=0.6$.}
\end{figure}
\thispagestyle{myheadings}
\begin{figure}[h!]
\figurenum{5(a)} \epsfysize=5.1875in {\vspace*{1.0in}
\epsffile{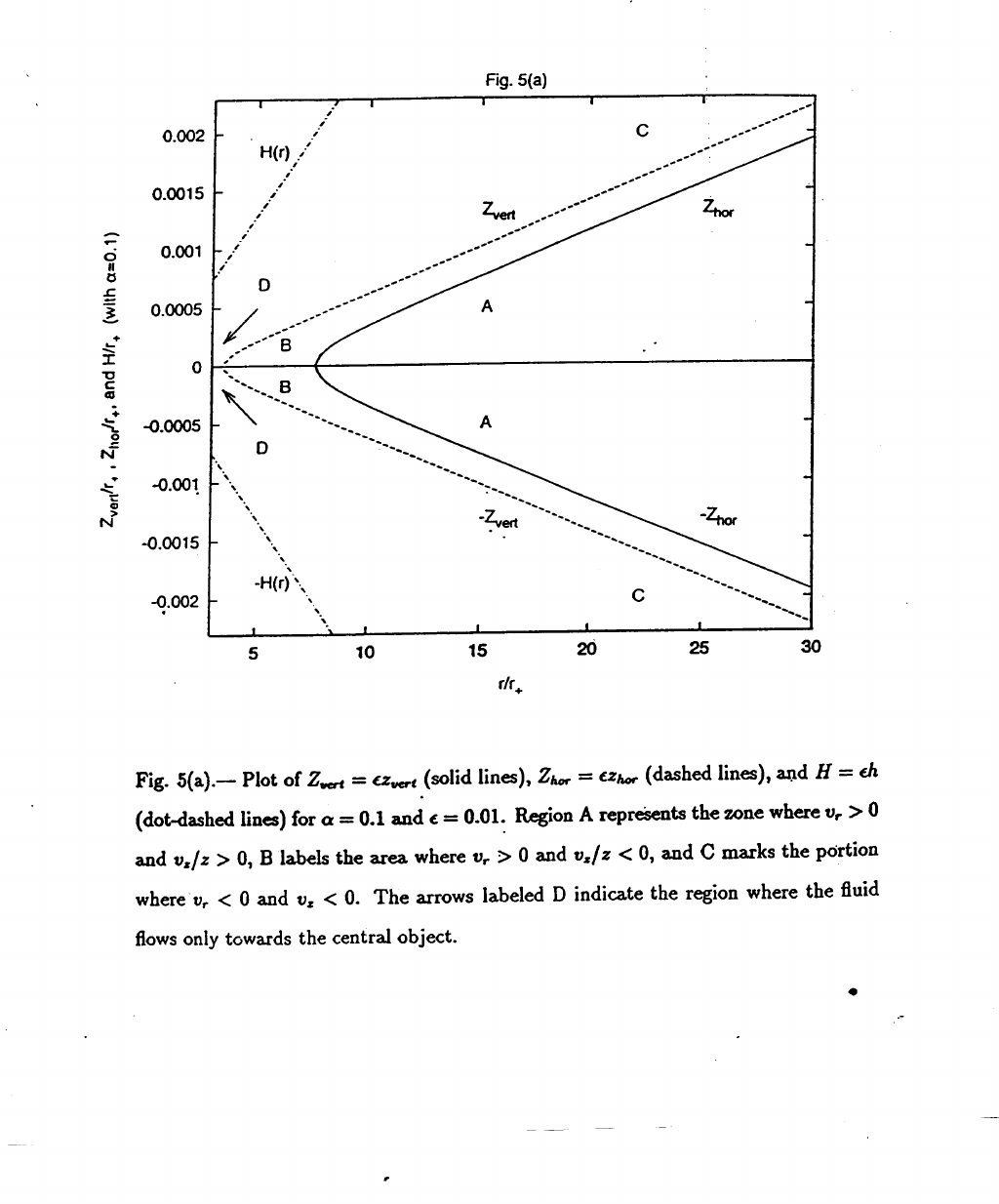}}
\caption{Plot of $Z_{vert}=\epsilon z_{vert}$ (solid lines),
$Z_{hor}=\epsilon z_{hor}$ (dashed lines), and $H=\epsilon h$
(dot-dashed lines) for $\alpha=0.1$ and $\epsilon=0.01$. Region A
represents the zone where $v_{r}>0$ and $v_{z}/z>0$, B labels the area
where $v_{r}>0$ and $v_{z}/z<0$, and C marks the portion where $v_{r}<0$
and $v_{z}<0$. The arrows labeled D indicate the region where the fluid flows
only towards the central object.}
\end{figure}
\thispagestyle{myheadings}
\begin{figure}[h!]
\figurenum{5(b)} \epsfysize=5.1875in {\vspace*{1.0in}
\epsffile{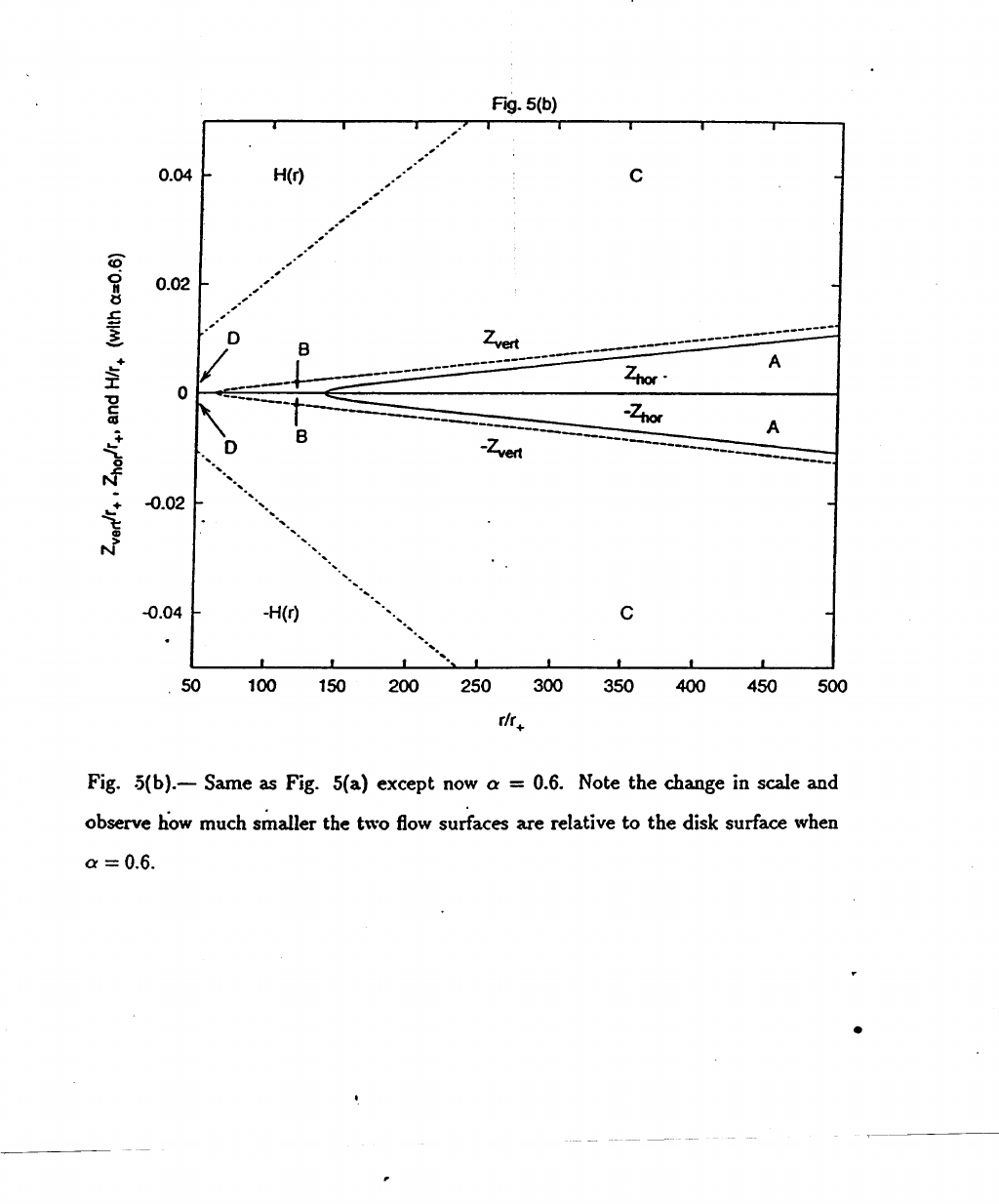}}
\caption{Same as Fig. 5(a) except now $\alpha=0.6$. Note the change in
scale and observe how much smaller the two flow surfaces are relative
to the disk surface when $\alpha=0.6$.}
\end{figure}

\thispagestyle{myheadings}
\begin{figure}[h!]
\figurenum{6(a)} \epsfysize=5.1875in {\vspace*{1.0in}
\epsffile{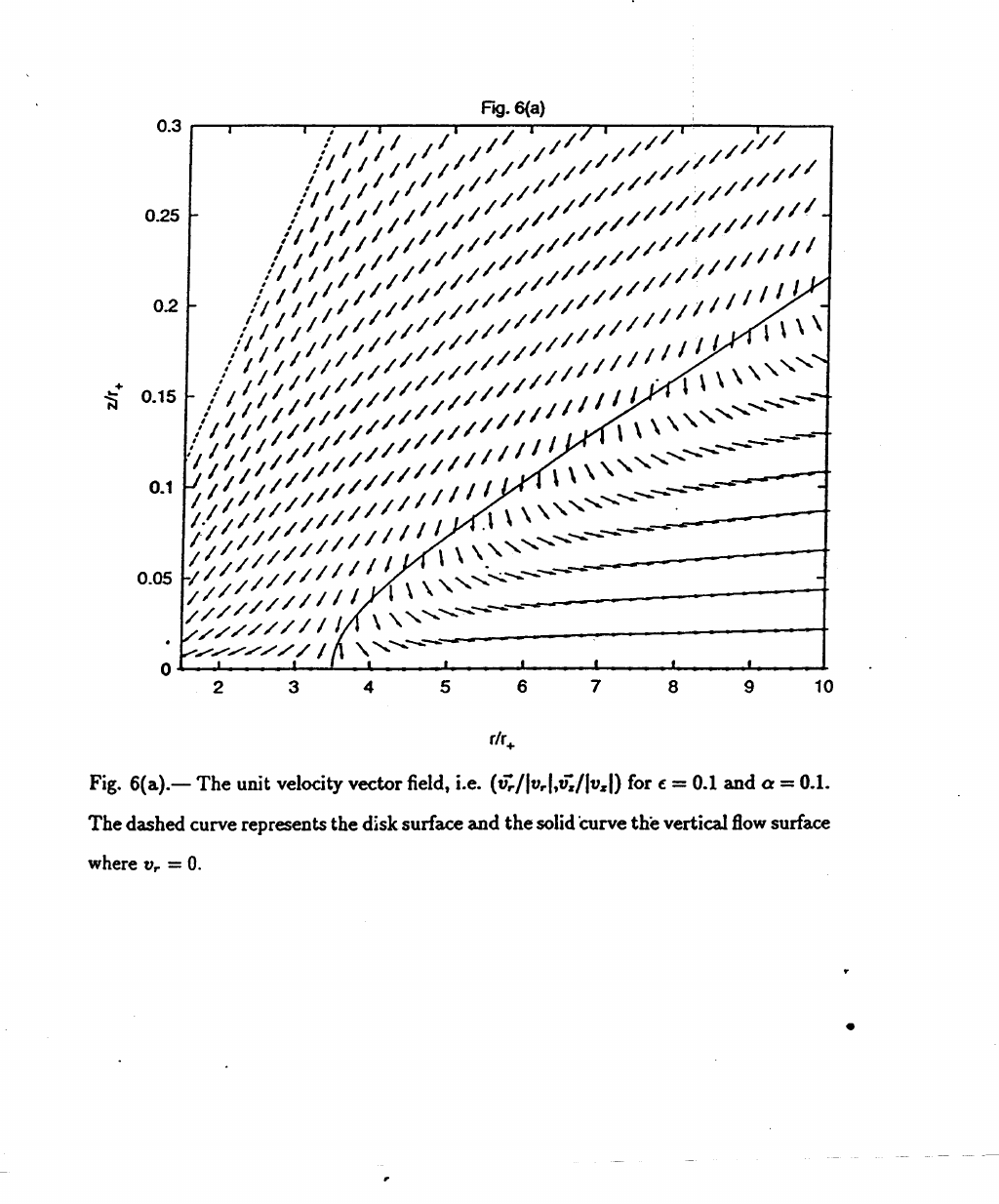}}
\caption{The unit velocity vector field, i.e. ($\vec{v_{r}}/\vert
v_{r} \vert$,$\vec{v_{z}}/\vert v_{z} \vert$) for $\epsilon=0.1$ and
$\alpha=0.1$. The dashed curve represents the disk surface and the
solid curve the vertical flow surface where $v_{r}=0$.}
\end{figure}
\thispagestyle{myheadings}
\begin{figure}[h!]
\figurenum{6(b)} \epsfysize=5.1875in {\vspace*{1.0in}
\epsffile{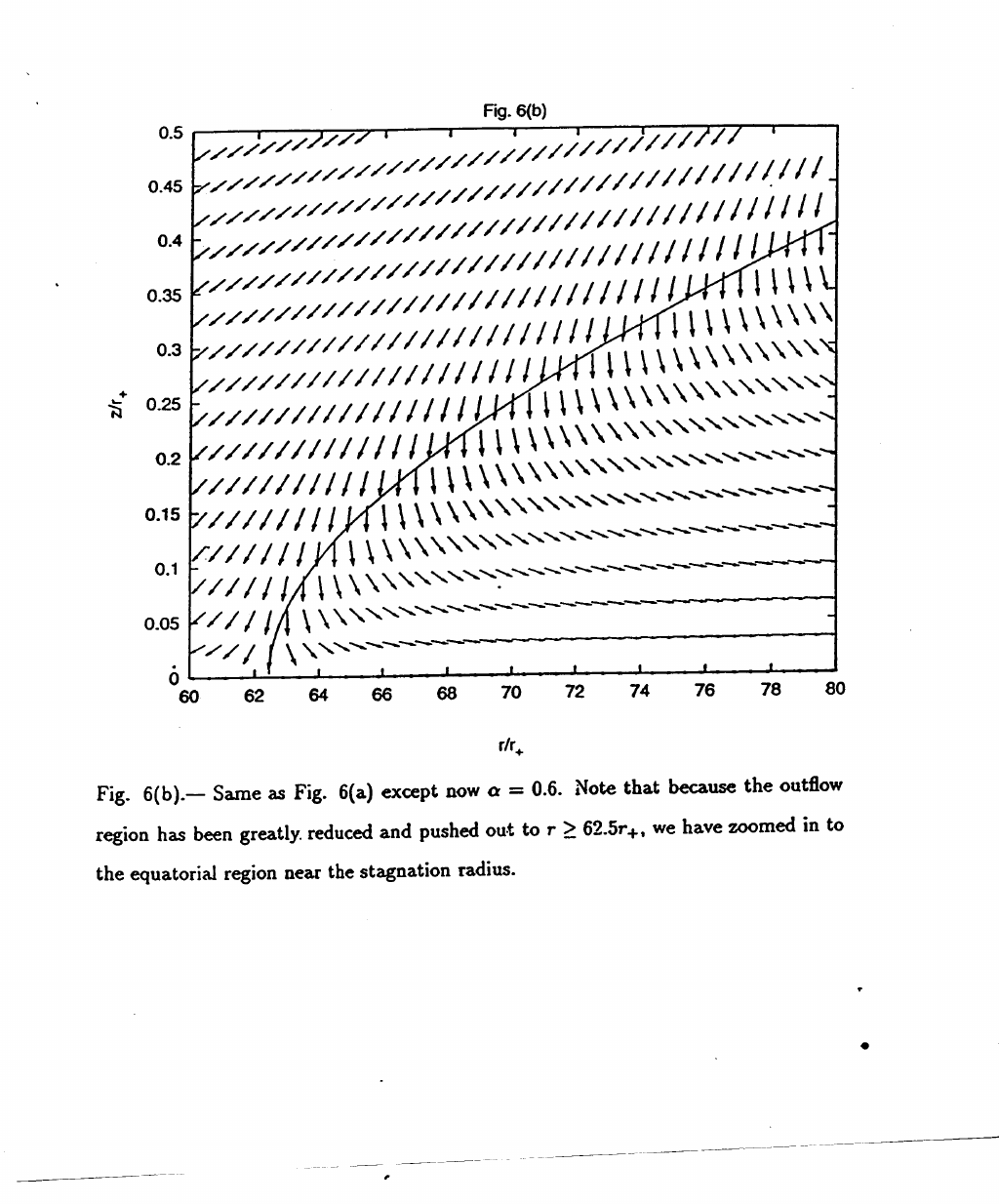}}
\caption{Same as Fig. 6(a) except now $\alpha=0.6$. Note that because
the outflow region has been greatly reduced and pushed out to $r\ge
62.5 r_{+}$, we have zoomed in to the equatorial region near the
stagnation radius.}
\end{figure}
\thispagestyle{myheadings}
\begin{figure}[h!]
\figurenum{6(c)} \epsfysize=5.1875in {\vspace*{1.0in}
\epsffile{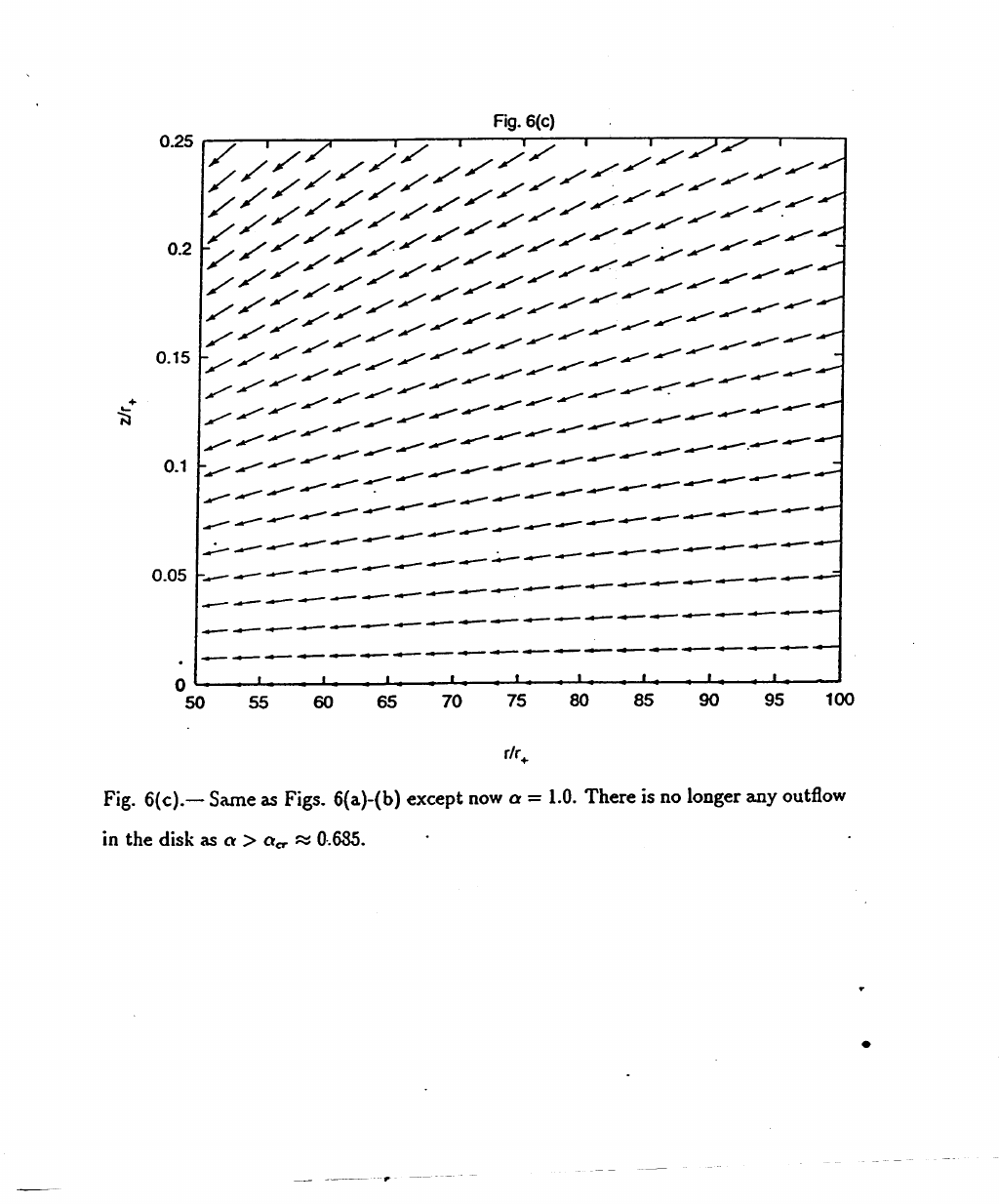}}
\caption{Same as Figs. 6(a)-(b) except now $\alpha=1.0$. There is no
longer any outflow in the disk as $\alpha > \alpha_{cr} \approx
0.685$.}
\end{figure}
\thispagestyle{myheadings}
\begin{figure}[h!]
\figurenum{7} \epsfysize=5.1875in {\vspace*{1.0in}
\epsffile{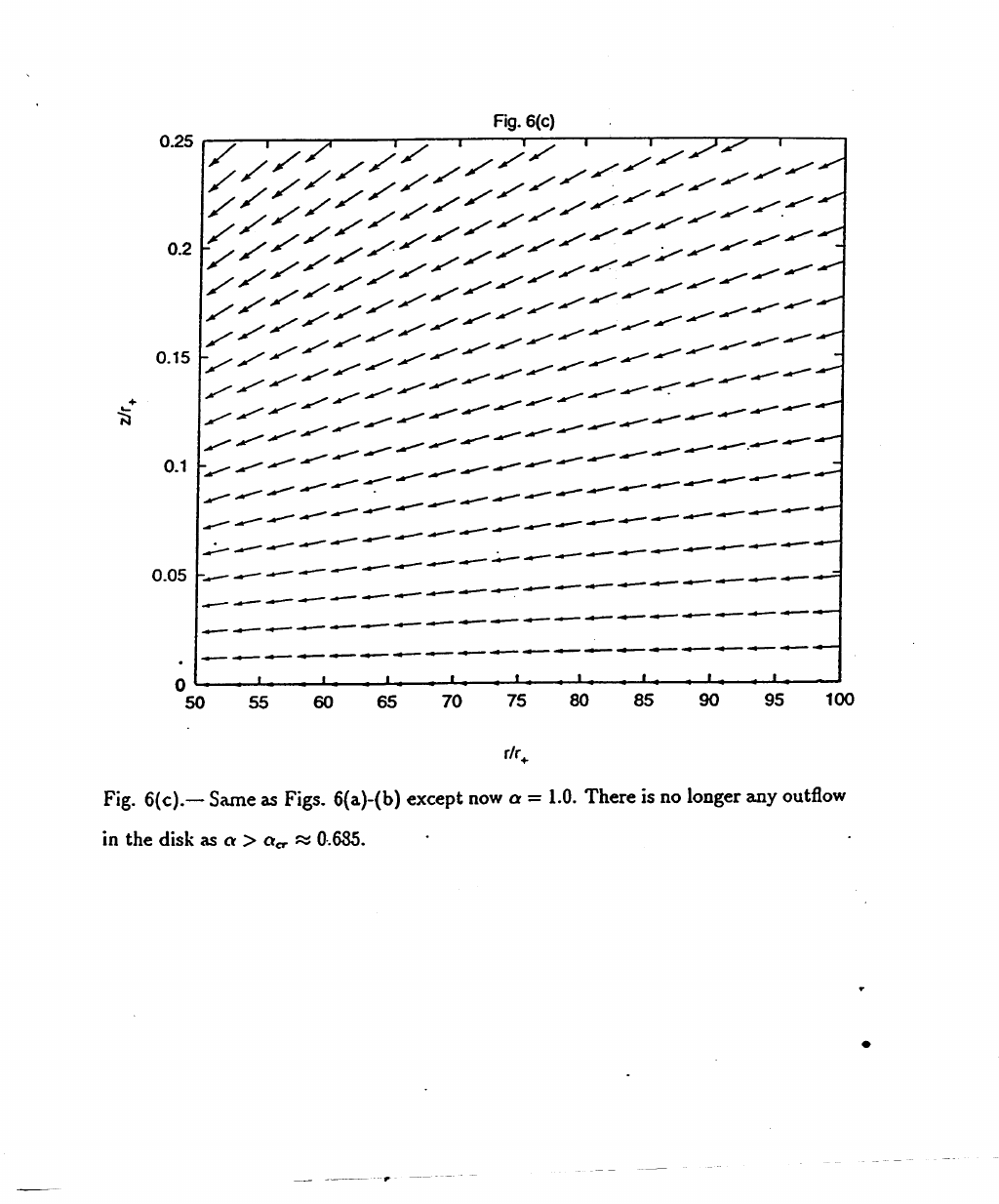}}
\caption{The ratio of the mass flow rate of material moving outwards
(through a cylinder with $r=\infty$) to the net accreted mass flow rate,
$\dot{M}$, vs. $\alpha$. Observe that $\Gamma_{\infty}(\alpha)$
vanishes as $\alpha \rightarrow \alpha_{cr}\approx 0.685$, but is
about $40\%$ for $\alpha<0.1$.}
\end{figure}

\end{document}